\documentclass[12pt]{article}
\usepackage{amsfonts, amsmath, amssymb, amsfonts, amscd, bm, booktabs, xcolor, enumitem, epstopdf, mathabx, hyperref, graphicx}
\usepackage[ruled]{algorithm2e}
\usepackage{dcolumn}
\newcolumntype{.}{D{.}{.}{-1}}
\newcolumntype{d}[1]{D{.}{.}{#1}}
\usepackage[top=.8in, left=.5in, right=.5in,
bottom=.8in]{geometry}
\RequirePackage{natbib}
\usepackage[compact]{titlesec}
\usepackage{booktabs}
\usepackage{url} 

\usepackage{color}

\newtheorem{remark}{Remark}

\newtheorem{theorem}{Theorem}
\newtheorem{corollary}{Corollary}

\newcommand{\sgn}{\mathrm{sgn}}

\newcommand{\bE}{\mathbb{E}}
\newcommand{\dto}{\xrightarrow[]{d}}

\usepackage{caption}
\usepackage{subcaption}

\expandafter\def\expandafter\normalsize\expandafter{\normalsize\setlength\abovedisplayskip{0pt}}
\expandafter\def\expandafter\normalsize\expandafter{\normalsize\setlength\belowdisplayskip{0pt}}
\expandafter\def\expandafter\normalsize\expandafter{\normalsize\setlength\abovedisplayshortskip{0pt}}
\expandafter\def\expandafter\normalsize\expandafter{\normalsize\setlength\abovedisplayshortskip{0pt}}

\begin{document}
\pagestyle{plain}

\def\spacingset#1{\renewcommand{\baselinestretch}%
{#1}\small\normalsize} \spacingset{1.5}


  \title{\bf Inference with non-differentiable surrogate loss in a general high-dimensional classification framework}
  \author{
  	Muxuan Liang\thanks{
  	Department of Biostatistics, The University of Texas MD Anderson Cancer Center}\\
  	\and
  	Yang Ning\thanks{
  	Department of Statistics and Data Science, Cornell University}\\
    \and
  	Maureen A Smith\thanks{
  	Departments of Population Health Sciences and Family Medicine
    \& Community Health, University of Wisconsin-Madison}\\
  	\and
  	Ying-Qi Zhao\thanks{
  	Public Health Sciences Division, Fred Hutchinson Cancer Center}}
  \date{}
  \maketitle

\thispagestyle{empty}

\abstract{
Penalized empirical risk minimization with a surrogate loss function is often used to learn a high-dimensional linear decision rule in classification problems. Although much of the literature focus on the generalization error, there is a lack of inference procedures for identifying the driving factors of the estimated decision rule, especially when the surrogate loss is non-differentiable. We propose a kernel-smoothed decorrelated score to construct hypothesis tests and interval estimators for a linear decision rule estimated using a piece-wise linear surrogate loss, which has a discontinuous gradient and non-regular Hessian. Specifically, we adopt kernel approximations to smooth the discontinuous gradient near discontinuity points and approximate the non-regular Hessian of the surrogate loss. In applications where additional nuisance parameters are involved, we propose a novel cross-fitted version to accommodate flexible nuisance estimates and kernel approximations. We establish the limiting distribution of the kernel-smoothed decorrelated score and its cross-fitted version in a high-dimensional setup. Simulation and real data analysis are conducted to demonstrate the validity and the superiority of the proposed method.
}

\newcommand{\n}{\noindent}
\maketitle

\newpage
\section{Introduction}
\label{sec:intro}

Classification identifies which of a set of labels an observation belongs to. Well-known classification methods include logistic regression, support vector machines, and many others. When the dimensionality of the covariate space is high, which is common due to the increasing adoption of large datasets in biomedical applications,  classification is more challenging \citep{fan2008high, dobriban2018high, bing2023optimal}. Additionally, it has been shown that many problems can be formulated within a general classification framework \citep{bartlett2006convexity, bartlett2008classification, Zhao2012, Zhou2017, zhao2019efficient}. In these problems, the goal is to derive a data-driven decision rule that minimizes a loss function (or maximizes a utility function) defined according to the problem setup. For example, in prediction problems, the goal is to derive a data-driven decision rule that minimizes the prediction error for binary outcomes or labels; in precision medicine, the goal is to derive a data-driven decision rule that maximizes the averaged utility over entire population if the derived decision rule were implemented to recommend treatment options \citep{Zhao2012, Zhou2017, zhao2019efficient}.

Empirical risk minimization (ERM) is often used to estimate such a data-driven decision rule by minimizing a convex surrogate of the loss function. Statistical inference of the constructed decision rules not only provides uncertainty quantification of the data-driven decision rule, but also enables a data-driven paradigm for new scientific discovery, e.g., to identify the risk factors for the outcome of interest in prediction problems; and to identify the driving factors of the estimated data-driven decision rule to inform new treatment guidelines in precision medicine \citep{jeng2018, wang2019distributed, ning2017, liang2022estimation}. However, while there is a large literature on classification and the generalization error of an ERM with a convex surrogate, statistical inference within a high-dimensional classification framework is less well-studied. For regression problems, \citet{vandegeer2014} proposed a debiased Lasso estimator for generalized linear models and established asymptotic normality under modest regularity conditions. \citet{ning2017} proposed a decorrelated score to test a low-dimensional projection of high-dimensional coefficient vectors, which can be applied to M-estimation under a strictly convex and differentiable loss function. \citet{chunhui2017} proposed a bootstrap procedure to conduct simultaneous inference for parameters in groups with diverging group sizes. More recently, partial penalized tests proposed in \citet{shi2019linear} can be applied to test hypotheses involving a growing number of coefficients. \citet{ma2020global} considered global hypothesis testing and multiple testing procedures for high-dimensional logistic regression models. \citet{wu2021model} proposed an inference procedure for single-index models with differentiable link functions. However, none of these methods can be applied to non-differentiable loss functions, such as the hinge loss, which is commonly adopted in classification problems.

Deriving an inference procedure for an ERM with a convex surrogate that is non-differentiable is more complicated. One such example is the popular classification method - support vector machine (SVM) \citep{cortes1995support}. It employs the hinge loss as a surrogate loss,  which is continuous but non-differentiable. The majority of the existing SVM literature focused on its consistency, and the convergence rate of the risk under the derived classifiers to the Bayes risk \citep{lin2004note, zhang2004statistical, steinwart2005consistency, zhang2004statistical, bartlett2006convexity, steinwart2007fast, vert2006consistency, blanchard2008statistical}. \citet{peng2016error} provided an error bound for a penalized SVM in ultra-high dimension; \citet{zhang2016consistent} and~\citet{zhang2016variable} focused on variable selection for SVMs in moderately high dimension. The literature on the asymptotic distribution of these estimators is limited. \citet{koo2008bahadur} investigated the Bahadur representation of a linear SVM, which implies the asymptotic normality of the estimator in a low-dimensional setup. Due to the non-differentiability of the hinge loss, they proposed a non-parametric estimator for the asymptotic variance. \citet{wang2019distributed} proposed a distributed inference procedure for a linear SVM. To handle the lack of differentiability, they used a smoothed loss function to approximate the hinge loss and showed the asymptotic normality of the estimator provided that $p/n\to 0$, where $n$ is the total sample size, and $p$ is the number of the covariates. However, an associated inference procedure in the high-dimensional setup is still lacking. 

Furthermore, the classification framework has been adapted to learn an individualized treatment rule (ITR), which recommends treatment according to patient characteristics. 
There has been much literature proposing to learn ITRs from a weighted classification framework, where the weights are related to the observed clinical outcomes \citep{Zhao2012, Zhou2017, chen2016personalized, yingqi2014, zhao2019efficient, pan2020improved, xue2020multicategory}. In recent work, \citet{zhao2019efficient} and \citet{liang2022estimation} introduced both the outcome regression models and propensity score as nuisance parameters in the weights. To avoid model misspecification, the nuisance parameters are estimated via nonparametric or flexible machine learning algorithms. These algorithms may lead to nuisance parameter estimators with slow convergence rates. These flexible estimators of the nuisance parameters with possible slow convergence rates create a large barrier in statistical inference. \citet{liang2022estimation} proposed an inference procedure that can handle strictly convex differentiable loss functions. However, inference procedure for non-differentiable loss functions remain largely unexplored.

We propose a novel inference procedure for linear decision rules under a general classification framework in a high-dimensional setup, which can deal with non-differentiable convex surrogate loss functions. We introduce a kernel-smoothed decorrelated score for the inference procedure, which utilizes a local kernel function to smooth the discontinuous gradient near discontinuity points, where the loss function is not differentiable, and a global kernel function to approximate the non-regular Hessian. 
By using these kernel functions, the proposed procedure can be applied to any piece-wise linear convex loss functions. Furthermore, unlike the existing literature \citep{wang2019distributed, koo2008bahadur}, the proposed procedure is valid even when $p/n\to +\infty$ and can be extended to test a hypothesis involving a growing number of projections. For the general classification problems, additional nuisance parameters may be involved in the loss function. Motivated by \citet{victor2018}, we further propose a new cross-fitting algorithm to efficiently accommodate these nuisance parameters. We show the uniform validity of the kernel-smoothed decorrelated score based procedure even when the nuisance parameters are estimated using nonparametric or flexible machine learning methods. Simulations and real data examples show the superiority of the proposed method.

Sections~\ref{sec:method_common} and~\ref{sec:class} introduce the kernel smoothed decorrelated score for the classification problem and the general classification framework. Section~\ref{sec:theory} provides theoretical justifications for the proposed procedure. 
Sections~\ref{sec:sim} and~\ref{sec:real_data} present simulations and real data analyses. Section~\ref{sec:diss} concludes the paper and provides a discussion on future directions.

\section{Statistical inference for classification problems}\label{sec:method_common}

\subsection{A classification problem}
\label{sec:common_class}


We observe the covariates $\bm X \in \mathbb{R}^p$ and a label $A\in \left\{-1,1\right\}$. A decision rule, $d:\mathbb{R}^p \rightarrow \left\lbrace -1, 1\right\rbrace$, is a  mapping from the covariate space $\mathbb{R}^p$ to the label space $\{-1,1\}$. We mainly focus on the high-dimensional setting with $p/n\to +\infty$. 
For any rule, $d$, define the classification error for $d$ as
\begin{eqnarray*}
	L(d)=\bE\left[1\left\{A\neq d(\bm X)\right\}\right],
	\label{classerror}
\end{eqnarray*}
which is also the expected zero-one loss to compare $d(\bm X)$ and $A$. The goal is to identify the decision rule which minimizes the classification error.  This optimal rule, also known as the Bayes rule, is given by 
$d_{\mathrm{opt}}(\mathbf{x}) = \mathrm{sgn}\left\lbrace 
P(A=1|\mathbf{X}=\mathbf{x}) - 1/2\right\rbrace$.

To estimate/infer this $d_{\mathrm{opt}}$, we observe both $\bm X$ and $A$. With these observed data, we optimize the empirical analogue of $L(d)$, denoted by $\widehat L(d) = \widehat{\bE}_n \left[1\left\{A\neq d(\bm X)\right\}\right] $, where $\widehat{\bE}_n[\cdot]$ is the empirical expectation. 
However, $\widehat L(d)$ is hard to optimize due to the discontinuity of the zero-one loss, and the space of decision rules is large. To tackle these challenges, a common strategy is to replace the zero-one loss by a piecewise convex loss function, often called a surrogate loss function, which is a real-valued map on $\mathbb{R}$. Examples of piece-wise convex loss functions include the hinge loss and the modified hinge loss used for classification with a rejection option \citep{bartlett2008classification}. Furthermore, we focus on linear decision rules. We thus optimize 
\begin{equation*}
	\widehat L_{\phi}(\bm\beta)=\widehat{\bE}_n [\phi(A\bm X^\top\bm \beta)],
\end{equation*}
where $\phi(\cdot)$ is a surrogate loss function. Define the minimizer of $L_{\phi}(\bm\beta)$ as $\bm\beta^*_{\phi}$, where $L_{\phi}(\bm\beta)= {\bE} [\phi(A\bm X^\top\bm \beta)]$. 
With high-dimensional data, we adopt penalized empirical risk minimization to estimate $\bm\beta^*_{\phi}$. Specifically, we consider 
\begin{equation*}
	\widehat{L}_{\phi}^{\lambda_n}(\bm\beta)=\widehat{\bE}_n[\phi(A\bm X^\top\bm \beta)]+\lambda_n\|\bm\beta\|_1,
\end{equation*}
where $\lambda_n$ is a tuning parameter. Denote the minimizer of $\widehat{L}_{\phi}^{\lambda_n}(\bm\beta)$ as $\widehat{\bm\beta}_{\phi}$. Through the estimator $\widehat{\bm\beta}_{\phi}$, we aim to construct a hypothesis testing procedure and confidence interval for a low-dimensional projection of $\bm\beta^*_{\phi}$, i.e. $\bm\eta^\top \bm\beta^*_{\phi}$, where $\bm\eta$ is a known sparse vector.

\begin{remark}\label{remark:norm}
In the classification problem, the classification error is determined by the linear direction represented by $\bm\beta^*_{\phi}$. Thus, when comparing with other methods, we normalize the estimates such that its first coordinate equals $1$, i.e., $|\beta_{\phi,1} ^*|=1$.
\end{remark}

\begin{remark}\label{remark:fisher}
Although hinge loss is Fisher consistent, the $\bm\beta_{\phi}^*$ does not necessarily equal the optimal linear rule minimizing $L(d)$. In classification problems, the hinge loss leads to the maximum-margin linear separator, which can be different from the 0–1-optimal linear hyperplane. This requires additional assumptions, e.g., sufficient conditions in ~\cite{liang2022estimation}; or a mixture of Gaussian model assumption (see Section~3 in \cite{koo2008bahadur}). We assume that the users have chosen to use hinge loss to derive a data-driven decision rule, and our focus is on how to conduct inference based on these estimates.
\end{remark}

\subsection{Kernel-smoothed decorrelated score}
\label{sec:score_common}

For the purpose of illustration, we consider a hypothesis testing problem $\mathcal{H}_0:\beta_{\phi,l}^*=0$ versus $\mathcal{H}_a:\beta_{\phi,l}^*\not = 0$, where $\beta^*_{\phi,l}$ is the $l$-th coordinate of the $\bm\beta_{\phi}^*$. The proposed approach can be easily applied to test $\mathcal{H}_0:\bm\eta^\top\bm\beta_{\phi}^*=0$ versus $\mathcal{H}_a:\bm\eta^\top\bm\beta_{\phi}^*\not = 0$.

We first review a decorrelated score, proposed in \citet{ning2017}, which can be used to test this hypothesis when the surrogate loss $\phi$ is differentiable. The key of the decorrelated score is to decouple the estimation error of the high-dimensional component $\bm\beta^*_{\phi, -l}$ from the estimation of $\beta_{\phi, l}^*$, where $\bm\beta^*_{\phi, -l}$ is the sub-vector of ${\bm\beta}_{\phi}^*$ without the $l$-th coordinate.   Define the minimizer of
\begin{equation*}
	L_{\phi^{''},l}\left (\bm w\right)={\bE}\left[\phi^{''}\left(A\bm X^\top{\bm\beta}^*_{\phi}\right)(X_l-\bm X_{-l}^\top\bm w)^2\right] 
\end{equation*}
as  $\bm w^*_{\phi,l}$.  The decorrelated score is defined as 
\begin{equation*}
	S_{\phi^{'},l}(\bm\beta; {\bm w}^*_{\phi,l})=\widehat{\bE}_n\left[A\phi^{'}\left(A\bm X^\top\bm\beta\right)(X_l-\bm X_{-l}^\top{\bm w}^*_{\phi,l})\right].
\end{equation*}
Let $\widehat{\bm\beta}_{\phi, null(l)}$ be a vector equals $\widehat{\bm\beta}_{\phi}$ except the $l$-th coordinate is fixed at $0$. Under the null hypothesis, it can be shown that
$
	\sqrt{n}S_{\phi^{'},l}(\widehat{\bm\beta}_{\phi, \mathrm{null}(l)}; {\bm w}^*_{\phi,l})\dto \mathcal{N}(0, \sigma_l^2),
$
where $\sigma^2_l$ is some constant. However, this procedure cannot be applied to the non-differentiable loss functions due to the non-existence of regular $\phi^{'}(\cdot)$ and $\phi^{''}(\cdot)$. 

For a piece-wise linear convex loss function $\phi$, although $\phi$ is not differentiable at the discontinuity points, the gradient is well defined on any open intervals without discontinuity points. Mathematically, suppose that $-\infty < t_1 < \cdots < t_J <+\infty$ are the jump discontinuity points of $\phi^{'}$, then $\phi^{'}$ can be defined as $\Delta_0+\sum_{j=1}^J \Delta_j1\left\{t-t_j\geq 0\right\}$ on any open intervals. In addition, a Hessian $\phi^{''}$ can be defined using the Dirac function $\delta(t)$ as $\sum_{j=1}^J \Delta_j\delta\left(t-t_j\right)$, which is non-regular because it  achieves $+\infty$ at $0$ and vanishes at all other points.

 We use kernel functions to smooth the gradient near discontinuity points and approximate the Hessian of the surrogate loss function. Specifically, we obtain a smoothed gradient $\phi^{'}$ by a \textbf{local} kernel function, a smooth function whose derivative has a support contained in a compact interval. Consider  $H(\cdot)$ satisfying $H(t)=1$ if $t\geq 1$ and $H(t)=0$ if $t\leq -1$; thus, its derivative has a support on $[-1,1]$. As the bandwidth $h_{\rm lo}\to 0$, the functions $H(t/h_{\rm lo})$ and $H'(t/h_{\rm lo})/h_{\rm lo}$ approach the indicator function $1(t\geq 0)$ and the Dirac function $\delta(t)$, respectively. One example of such kernel is
\begin{equation*}
	H(t)=
	\begin{cases}
		0 & \text{if } t\leq -1,\\
		\frac{1}{2}+\frac{15}{16}(t-\frac{2}{3}t^3+\frac{1}{5}t^5) &\text{if } |t|<1,\\
		1 & \text{if } t\geq 1.
	\end{cases}
\end{equation*}
Define
\begin{equation*}
	\widetilde{\phi}^{'}(t)= \Delta_0+\sum_{j=1}^J \Delta_j H\left(\frac{t-t_j}{h_{\rm lo}}\right),
\end{equation*}
where $\Delta_j=\phi^{'}(t_j+)-\phi^{'}(t_j-)$. 
For any open interval where $\phi^{'}$ exists, $\widetilde{\phi}^{'}(t)$ is different from $\phi^{'}$ only on $\bigcup_{j=1}^J [t_j-h_{\rm lo}, t_j+h_{\rm lo}]$. 
Thus, the smoothed gradient of $L_{\phi}(\bm\beta)$ can be naturally defined by
$
	{\bE}\left[A\widetilde{\phi}^{'}\left(A\bm X^\top{\bm\beta}\right)\bm X\right]. 
$ 
The smoothed score function of $\beta_{\phi,l}^*$ is
$
{\bE}\left[A\widetilde{\phi}^{'}\left(A\bm X^\top\widehat{\bm\beta}_{\phi, \rm null(l)}\right)X_l\right]. 
$
However, in high dimension, it is biased due to the estimation error of the high-dimensional component $\widehat{\bm\beta}_{\phi,-l}$. Following the idea of the decorrelated score, we decouple the smoothed score function of $\beta_{\phi,l}^*$ with the estimation error of $\widehat{\bm\beta}_{\phi,-l}$. Define $\bm w^*_{\phi,l}$ as the minimizer of
\begin{equation*}
	L_{\phi^{''},l}\left (\bm w\right)={\bE}\left[\sum_{j=1}^J{\Delta_j}\delta(t_j-A\bm X^\top{\bm\beta}^*_{\phi})(X_l-\bm X_{-l}^\top\bm w)^2\right].
\end{equation*}
We construct the kernel-smoothed decorrelated score 
\begin{equation*}
	S_{\widetilde{\phi}^{'},l}(\bm\beta; {\bm w}_{\phi,l}^*)=\widehat{\bE}_n\left[A\widetilde{\phi}^{'}\left(A\bm X^\top{\bm\beta}\right)(X_l-\bm X_{-l}^\top {\bm w}_{\phi,l}^*)\right].
\end{equation*}

To construct a score test for $\beta_{\phi,l}^* = 0$ using $S_{\widetilde{\phi}^{'},l}(\bm\beta; {\bm w}_{\phi,l}^*)$, we need to estimate $\bm w^*_{\phi,l}$, which is nontrivial due to the non-regularity of the $\delta(t)$ function. We propose to use a \textbf{global} kernel function with a support on the entire real line to approximate $\delta(t)$. Suppose that $G(t)$ is a global kernel function satisfying that 1) $G(t)>0$, $\forall t$; 2) $\int G(t)\textrm{d}t=1$; 3) $\int tG(t)\textrm{d}t=0$. Define $G_{h_{\rm gb}}(t)=h_{\rm gb}^{-1}G(t/h_{\rm gb})$, where $h_{\rm gb}$ is the bandwidth of the global kernel. The estimated $\bm w^*_{\phi,l}$, denoted by $\widehat{\bm w}_{\phi,l}$, can be obtained by minimizing 
\begin{equation*}
	\widehat{\bE}_n\left[\sum_{j=1}^J \Delta_j G_{h_{\rm gb}}(t_j-A\bm X^\top\widehat{\bm\beta}_{\phi})(X_l-\bm X_{-l}^{\top}\bm w)^2\right]+\mu_n\|\bm w\|_1,
\end{equation*}
where $\mu_n$ is a tuning parameter. The above objective function is strictly convex and smooth with respect to $\bm w$ given that $G(t)>0$, and thus can be easily minimized.  

\begin{remark}
A straightforward idea is to use the gradient of the local kernel to approximate $\delta(t)$. However, only data points near the discontinuity points will contribute to the estimation of $\bm w^*_{\phi,l}$ in this case, because further derivative of $\tilde{\phi}^{'}(t)$ is zero when $t$ is away from the discontinuity points. 
In supplementary material, we show that using the local kernel function to approximate the non-regular Hessian will lead to a slower convergence rate of the estimator and require more restrictive conditions in the inference procedure.
\end{remark}
\begin{remark}
    Kernel functions are used to smoothen the gradient and hessian around non-differentiable points, which lead to gradient and hessian that can be easily computed. For non-differentiable loss, it may be possible to use sub-gradient of $\phi$ to replace $\phi^{'}(\cdot)$ and sub-hessian of $\phi$ to replace $\phi^{''}(\cdot)$; however, this approximation may be computationally challenging and problematic. For example, hinge loss has vanishing (sub-)hessian almost everywhere, and thus a more detailed approximation, e.g., using a global/local kernel function, is needed.
\end{remark}

Replacing the $\bm w^*_{\phi,l}$ by $\widehat{\bm w}_{\phi,l}$ in $S_{\widetilde{\phi}^{'},l}(\bm\beta; {\bm w}_{\phi,l}^*)$, we can calculate the value of the estimated decorrelated score function $ S_{\widetilde{\phi}^{'},l}(\widehat{\bm\beta}_{\phi, null(l)}; \widehat{\bm w}_{\phi,l})$ to test the
 null hypothesis. However, additional challenges arise from the correlation between $\widehat{\bm\beta}_{\phi}$ and the loss function used to estimate $\bm w^*_{\phi,l}$, as well as the correlation between $\widehat{\bm\beta}_{\phi}$ and the kernel-smoothed decorrelated score function $S_{\widetilde{\phi}^{'},l}(\cdot)$. We design a novel sample-splitting strategy, where the estimator $\widehat{\bm\beta}_{\phi}$ is independent from the data used to estimate $\bm w^*_{\phi,l}$, as well as the data used to construct $S_{\widetilde{\phi}^{'},l}(\cdot)$. In addition, instead of averaging over multiple estimators as in the cross-fitting procedure \citep{victor2018}, we average over the loss functions to estimate $\widehat{\bm w}_{\phi,l}$, which is more computationally robust. The entire inference procedure is summarized in Algorithm~\ref{algorithm:1}. In our simulation and real data analysis, we set $K=2$. For the bandwidth selection, we choose  $h_{\rm lo}=1/\sqrt{n\log n}$ and $h_{\rm gb}=(\log p/n)^{-1/5}$ based on the theoretical results in Section~\ref{sec:thoery_wo_nusiance} and~\ref{sec:thoery_w_nusiance}.

\begin{algorithm}
		\caption{Inference of $\bm\beta^*_{\phi}$.}
		\label{algorithm:1}
		\SetAlgoLined
		\KwIn{A random seed; $n$ samples; a positive integer $K$.}
		\KwOut{A p-value for $\mathcal{H}_0:\beta^*_{\phi,l}=0$.}
		\label{algo1:1} Randomly split data into $K$ parts $I_1,\cdots, I_K$ with equal size, and set $k=1$\;
		\label{algo1:2} Estimate $\bm\beta^*_{\phi}$ using data in ${I}_{k}^c$ by
		\begin{equation*}
			\widehat{\bE}_n^{(-k)}[\phi(A\bm X^\top\bm \beta)]+\lambda_n^{(-k)}\|\bm\beta\|_1,
		\end{equation*} 
		where $\widehat{\bE}_n^{(-k)}[\cdot]$ is the empirical average on $I_k^c$, and denote the estimator as $\widehat{\bm\beta}^{(-k)}_{\phi}$. The parameter $\lambda_n^{(-k)}$ is tuned by cross-validation; we obtain $\widehat{\bm\beta}^{(-k)}_{\phi}$ for each $k$\;
		\label{algo1:4} Obtain an estimator $\widehat{\bm w}_{\phi,l}$ for $ \bm w^*_{\phi,l} $by minimizing
		\begin{eqnarray*}
			\frac{1}{K}\sum_{k=1}^K\widehat{\bE}_n^{(k)}\left[\left\{\sum_{j=1}^J \Delta_j G_{h_{\rm gb}}(t_j-A\bm X^\top\widehat{\bm\beta}^{(-k)}_{\phi})\right\}(X_l-\bm X_{-l}^{\top}\bm w)^2\right]+\mu_n\|\bm w\|_1,
		\end{eqnarray*}
		where $\mu_n$ is tuned by cross-validation, and $\widehat{\bE}_n^{(k)}[\cdot]$ is the empirical average on $I_k$\;
		\label{algo1:5}  Let $\left(\widehat{\bm \beta}_{\phi, null(l)}^{(k)}\right)^{\top}$ equal to $\widehat{\bm\beta}_{\phi}^{(k)}$ except its $l$-th coordinate replaced by $0$. Construct the kernel-smoothed decorrelated score test statistic as
		$$
		S_{\widetilde{\phi}^{'}, null(l)}  = \frac{1}{K}\sum_{k=1}^K\widehat{\bE}_n^{(k)}\left[A\widetilde{\phi}^{'}\left(A\bm X^\top\widehat{\bm\beta}_{\phi, null(l)}^{(-k)}\right)(X_l-\bm X_{-l}^\top\widehat{\bm w}_{\phi,l})\right],
		$$ 
		and the estimator of the variance $\widehat{\sigma}_l^2=\frac{1}{K}\sum_{k=1}^K\widehat{\bE}_n^{(k)}\left[\left\{\widetilde{\phi}^{'}\left(A\bm X^\top\widehat{\bm\beta}_{\phi}^{(-k)}\right)(X_l-\bm X_{-l}^\top\widehat{\bm w}_{\phi,l})\right\}^2\right]$\;
		\label{algo1:6} 
		Calculate the p-value by $2\left(1-\Phi(n^{-1/2}|S_{\widetilde{\phi}^{'},null(l)}|/\widehat{\sigma}_l)\right)$, where $\Phi(\cdot)$ is the cumulative distribution function of a standard normal distribution. 
	\end{algorithm}
	
\subsection{Construction of confidence interval}
\label{sec:ci}

Due to the penalization, the estimator $\widehat{\beta}_{\phi, l}$ is biased and cannot be directly used to construct an interval estimator. To remove this bias, the key idea is to consider a one-step estimator based on an unbiased estimating equation. Motivated by the fact that the kernel-smoothed decorrelated score is an asymptotically unbiased estimating equation for $\beta_{\phi,l}^*$, when the bandwidth shrinks to 0, we consider
\begin{equation*}
	\widetilde{\beta}_{\phi, l}=\bar{\beta}_{\phi, l}-S_{\tilde{\phi}^{'},l}(\widehat{\bm\beta}_{\phi}^{(-k)}; \widehat{\bm w}_{\phi,l})/\widehat{I}_l,
\end{equation*}
where 
\begin{eqnarray*}
    \bar{\beta}_{\phi, l}
    &=&\frac{1}{K}\sum_{k=1}^K \widehat{\beta}_{\phi, l}^{(-k)},\\
	S_{\tilde{\phi}^{'},l}(\bm\beta; \widehat{\bm w}_{\phi,l})
	&=&\frac{1}{K}\sum_{k=1}^K\widehat{\bE}_n^{(k)}\left[A\widetilde{\phi}^{'}\left(A\bm X^\top{\bm\beta}\right)(X_l-\bm X_{-l}^\top\widehat{\bm w}_{\phi,l})\right],\\
	\widehat{I}_l
	&=&\frac{1}{K}\sum_{k=1}^K\widehat{\bE}_n^{(k)}\left[\sum_j \Delta_j G_{h_{\rm gb}}\left(t_j-A\bm X^\top\widehat{\bm\beta}_{\phi}^{(-k)}\right)(X_l-\bm X_{-l}^\top\widehat{\bm w}_{\phi,l})^2\right],
\end{eqnarray*}
can be calculated in Step~\ref{algo1:6} of Algorithm~\ref{algorithm:1}. Then $95\%$-confidence interval can be constructed by $$\left(\widetilde{\beta}_{\phi, l}-1.96n^{-1/2}\widehat{\sigma}_l/\widehat{I}_l, \widetilde{\beta}_{\phi, l}+1.96n^{-1/2}\widehat{\sigma}_l/\widehat{I}_l\right).$$
	
\section{A general classification framework}\label{sec:class}

Weighted classification assigns class-specific weights, which reflect the relative importance of each class. More broadly, weights can depend on the covariate $\bm X$ that each individual has specific weights. Mathematically, the task is to learn a decision rule, $d(\bm X)$, which minimizes a weighted  zero-one loss 
\begin{eqnarray}\label{eq:zero_one_loss}
	&&L(d; W_1, W_{-1})=\bE[W_1(\bm X)1\left\{d(\bm X)\neq1\right\}+W_{-1}(\bm X)1\left\{d(\bm X)\neq-1\right\}],
\end{eqnarray}
where $W_a(\bm X)$'s are pre-specified weights depending on the problem of interest. 

Under the proposed general classification framework, the minimizer of the loss~\eqref{eq:zero_one_loss} can be derived as 
\begin{equation*}
	d_{\rm opt}(\bm X)=\sgn\left\{W_1(\bm X)-W_{-1}(\bm X)\right\},
\end{equation*}
where $\sgn(t)=1$ if $t\geq 0$; $\sgn(t)=-1$, otherwise.

We are interested in developing an inference procedure for a low-dimensional projection of $\bm\beta^*_{\phi}$, where  $\bm\beta^*_{\phi}$  is the minimizer  of
\begin{eqnarray*}
	L_{\phi}(\bm\beta; W_1, W_{-1})=\bE\left[W_1 \phi(\bm X^\top\bm\beta) +W_{-1}\phi(-\bm X^\top\bm\beta)\right],
\end{eqnarray*}
where for simplicity we use the notation $W_1$ and $W_{-1}$. The weights are typically not directly observed, and need to be estimated. Specifically, to estimate $\bm\beta^*_{\phi}$, we minimize  
\begin{eqnarray*}
	&&\widehat L_\phi^{\lambda_n}(\bm\beta;  \widehat{W}_1, \widehat{W}_{-1})=\widehat{\bE}_n\left[\widehat{W}_1\phi(\bm X^\top\bm\beta)+\widehat{W}_{-1}\phi(-\bm X^\top\bm\beta)\right]+\lambda_n\|\bm\beta\|_1,
\end{eqnarray*}
where $\widehat{W}_a$'s are the estimated weights using the observed data. 

The general weighted classification framework has been broadly used in many applications.  We provide two examples below. 

\subsection{Classification with missing labels} 
In a classification problem, it is likely that only partial samples are fully observed with $(\bm X, A)$, and  we only observe the covariate information $\bm X$ for the remaining samples. For example, to predict patient-reported outcomes, covariate information is collected at baseline, whereas the outcomes are collected by a survey after intervention. We can only observe outcomes for those patients who fill out the survey, and other patients' outcomes are missing. This problem is also related to the semi-supervised learning literature, where the missing is often assumed to be completely at random \citep{wang2007large, hoffmann2020consistency, song2024general, deng2024optimal, cai2025semi}. In this case, our proposed method can also be applied to infer the derived linear rule in semi-supervised learning.

Let $R$ be the missing indicator. We assume  missing at random (MAR),   i.e., $A\perp R\mid \bm X$.  By leveraging both the labeled and unlabeled samples, an estimator of the classification error is 
\begin{eqnarray}
&&\widehat{\bE}_n\left[\widehat{W}_1(\bm X, A;\widehat{\pi}, \widehat{p}) 1\{d(\bm X) \neq 1\}+\widehat{W}_{-1}(\bm X, A;\widehat{\pi}, \widehat{p}) 1\{d(\bm X) \neq -1\}\right].
\label{errorestimate}
\end{eqnarray}
Here,
\begin{eqnarray*}
\widehat{W}_a(\bm X, A;\widehat{\pi}, \widehat{p})=\frac{1\{R=1, A=a\}}{\widehat{\pi}_1(\bm X)}- \frac{1\{R=1\}-\widehat{\pi}_1(\bm X)}{\widehat{\pi}_1(\bm X)} \widehat{p}_a(\bm X),
\end{eqnarray*}
where $\widehat{\pi}_1(\bm X)$ is an estimate of the nuisance parameter $\pi_1=P(R=1\mid \bm X)$ and $\widehat{p}_a(\bm X)$ is an estimate of the nuisance parameter $p_a(\bm X) =  P(A=a|\bm X)$. Let $\widebar{\pi}_1(\bm x)$ and $\widebar{p}_a(\bm x)$ be the point-wise limits of the $\widehat{\pi}_1(\bm x)$ and $\widehat{p}_a(\bm x)$. Define
\begin{eqnarray*}
\widebar{W}_a(\bm X, A;\widebar{\pi}, \widebar{p})=\frac{1\{R=1, A=a\}}{\widebar{\pi}_1(\bm X)}- \frac{1\{R=1\}-\widebar{\pi}_1(\bm X)}{\widebar{\pi}_1(\bm X)} \widebar{p}_a(\bm X).
\end{eqnarray*}
 Under the class of linear decision rules, we can minimize 
\begin{eqnarray*}
&&\widehat{L}_{\phi}^{\lambda_n}(\bm\beta; \widehat{W}_1, \widehat{W}_{-1})=\widehat{\bE}_n\left[\widehat{W}_1(\bm X, A;\widehat{\pi}, \widehat{p}) \phi(\bm X^\top\bm\beta)+ \widehat{W}_{-1}(\bm X, A;\widehat{\pi}, \widehat{p}) \phi(-\bm X^\top\bm\beta)\right]+\lambda_n\|\bm\beta\|_1.
\end{eqnarray*}
We assume that there exists a constant $c$ such that $0<c < \min\{\pi_1, \widehat{\pi}_1\}$. The target of the inference procedure is a low-dimensional projection of $\bm\beta_{\phi}^*$, which minimizes
\begin{eqnarray}
&&L_{\phi}(\bm\beta; \widebar{W}_{1}, \widebar{W}_{-1})=\bE\left[\widebar{W}_1(\bm X, A;\widebar{\pi}, \widebar{p}) \phi(\bm X^\top\bm\beta) +\widebar{W}_{-1}(\bm X, A;\widebar{\pi}, \widebar{p}) \phi(-\bm X^\top\bm\beta)\right].\label{eq:loss_w_bar}
\end{eqnarray}
Further, due to the construction of the $\widebar{W}_a(\bm X, A;\widebar{\pi}, \widebar{p})$'s, if either $\widebar{\pi}_1=\pi_1$ or $\widebar{p}_a=p_a$, then $\bE[\widebar{W}_a\mid \bm X]=W_a(\bm X)\equiv P(A=a|\bm X)$ and the objective function~\eqref{eq:loss_w_bar} equals to
\begin{eqnarray*}
L_{\phi}(\bm\beta; {W}_{1}, {W}_{-1})=\bE\left[{W}_1\phi(\bm X^\top\bm\beta) +{W}_{-1} \phi(-\bm X^\top\bm\beta)\right].
\end{eqnarray*}

\subsection{Estimation of individualized treatment rules} 
An individualized treatment rule $d(\bm X)$ maps the covariate space $\mathbb{R}^p$ to the treatment space $\{-1, 1\}$, which is the label space here. To define the objective function,  we adopt the potential outcome framework \citep{rubin1974, rubin2005}. Denote the potential outcome under treatment $a\in\{-1, 1\}$ as $Y(a)$, and the potential outcome under an individualized treatment rule $d$ as $Y(d)$. Assume larger outcomes are more preferable. The goal is to learn the optimal individualized treatment rule that maximizes $\bE\left[Y(d)\right]$. 

 We observe the covariate information $\bm X$, the assigned treatment $A$, and the outcome $Y$. We assume the following conditions: 1) the Stable Unit Treatment Value Assumption (SUTVA) \citep{imbens_rubin_2015}; 2) the strong ignorability $Y(-1),Y(1)\perp A\mid \bm X$; 3) Consistency $Y = Y(a)$ if $A=a$. 
 SUTVA condition assumes that the potential outcomes for a patient do not vary with the other patients' treatments. It also implies that there are no different versions of the treatment. The strong ignorability condition means that there is no unmeasured confounding between the potential outcomes and the treatment. The consistency ensures that the observed outcome is the potential outcome under the assigned treatment. Under these conditions, an augmented inverse probability weighted estimator of $\bE\left[Y(d)\right]$ is 
\begin{eqnarray}
 &&\widehat{\bE}_n\left[\widehat{W}_1(Y, \bm X, A; \widehat{p}, \widehat{Q}) 1\{d(\bm X) = 1\}+\widehat{W}_{-1}(Y, \bm X, A; \widehat{p}, \widehat{Q})1\{d(\bm X) = -1\}\right].
\label{vestimate}
\end{eqnarray}
Here, 
\begin{eqnarray*}
&&\widehat{W}_a(Y, \bm X, A; \widehat{p}, \widehat{Q})=\frac{Y1\left\{A=a\right\}}{\widehat{p}_a(\bm X)}+\frac{1\left\{A=a\right\}-\widehat{p}_a(\bm X)}{\widehat{p}_a(\bm X)}\widehat{Q}_a(\bm X),
\end{eqnarray*}
where $\widehat{p}_a(\bm X)$ and $\widehat{Q}_a(\bm X)$ are estimators for the nuisance parameters $p_a(\bm X)=P(A=a\mid \bm X)$ and ${Q}_a(\bm X) = \bE(Y|\bm X, A = a)$, respectively. We assume that there exists a constant $c>0$ such that $c< p_a(\bm X), \widehat{p}_a(\bm X)<1-c$.  
 Let $\widebar{p}_a$ and $\widebar{Q}_a$ be the point-wise limit of $\widehat{p}_a$ and $\widehat{Q}_a$. Define
\begin{eqnarray*}
\widebar{W}_a(Y, \bm X, A;\widebar{p}, \widebar{Q})=\frac{Y1\left\{A=a\right\}}{\widebar{p}_a(\bm X)}+\frac{1\left\{A=a\right\}-\widebar{p}_a(\bm X)}{\widebar{p}_a(\bm X)}\widebar{Q}_a(\bm X).
\end{eqnarray*}

Consider the class of linear decision rules, we can minimize 
\begin{eqnarray}\label{eq:loss_w_bar_itr}
	&&\widehat{L}_{\phi}^{\lambda_n}(\bm\beta; \widehat{W}_1, \widehat{W}_{-1})=\widehat{\bE}_n\left[\sum_{a\in\{-1,1\}}\widehat{W}_a(Y, \bm X, A; \widehat{p}, \widehat{Q})\phi(\bm aX^\top\bm\beta)\right]+\lambda_n\|\bm\beta\|_1.
\end{eqnarray}

The target of the inference procedure is a low-dimensional projection of $\bm\beta^*_{\phi}$ defined as the minimizer of 
\begin{eqnarray*}
	&&L_{\phi}(\bm\beta;\widebar{W}_{1}, \widebar{W}_{-1})
=\bE\left[\widebar{W}_1(\bm X, A;\widebar{p}, \widebar{Q}) \phi(\bm X^\top\bm\beta) +\widebar{W}_{-1}(\bm X, A;\widebar{p}, \widebar{Q}) \phi(-\bm X^\top\bm\beta)\right].
\end{eqnarray*}

Furthermore, if either $\widebar{p}_a(\bm X) = p_a(\bm X)$ or $\widebar{Q}_a(\bm X) = {Q}_a(\bm X)$, then $\bE[\widebar{W}_a\mid \bm X]=W_a(\bm X)\equiv Q_a(\bm X)$, and the above objective function is equivalent to
\begin{eqnarray*}
L_{\phi}(\bm\beta; W_{1}, W_{-1})=\bE\left[W_1\phi(\bm X^\top\bm\beta) +{W}_{-1} \phi(-\bm X^\top\bm\beta)\right].
\end{eqnarray*}

In the general classification framework where nuisance parameters are involved, nonparametric or machine learning algorithms are commonly employed to fit them to avoid model misspecification. However, the convergence rates of the $\widehat{W}_a$'s may be slower than $O_p(n^{-1/2})$.  We adopt a cross-fitting procedure \citep{chernozhukov2017double} to tackle this issue. We split the entire dataset into two halves. The first half is used to fit the nuisance parameters, and  to estimate the weights,  $\widehat{W}_a$. We then implement Algorithm~\ref{algorithm:1} on the second half of the data to obtain the estimated coefficients and p-values. Similarly, we can then fit the nuisance parameters on the second half and use the first half to estimate the coefficients and conduct inference. Finally, to compensate for the efficiency loss due to the cross-splitting, we can average the estimates and the kernel-smoothed decorrelated scores. The details of this algorithm can be found in Algorithm~\ref{algorithm:2}.

\begin{algorithm}
		\caption{Inference of $\bm\beta^*_{\phi, 1}$ with nuisance parameters.}
		\label{algorithm:2}
		\SetAlgoLined
		\KwIn{A random seed; $n$ samples; a positive integer $K$.}
		\KwOut{A p-value for $\mathcal{H}_0:\beta^*_{\phi,l}=0$.}
		\label{algo2:1} Randomly split data into halves $\widetilde{I}$ and $\widetilde{J}$ with equal sizes\;
		\label{algo2:2} Estimate nuisance parameters using data in $\widetilde{I}$ by kernel regression after variable screening, and construct the estimated weights $\widehat{W}_a^{(\widetilde{I})}$'s on the samples in $\widetilde{J}$ using the estimated nuisance parameters\;
		\label{algo2:3} On $\widetilde{J}$, we implement Algorithm~\ref{algorithm:1} with weights $\widehat{W}_a^{(\widetilde{I})}$'s and denote the kernel-smoothed decorrelated score and its variance estimate as $S_{\widetilde{\phi}^{'}, null(l)}^{(\widetilde{J})}$ and $\widehat{\sigma}^2_{(\widetilde{J}),l}$\;
		\label{algo2:4} Similarly, we can obtain $S_{\widetilde{\phi}^{'}, null(l)}^{(\widetilde{I})}$ and $\widehat{\sigma}^2_{(\widetilde{I}),l}$.
		 Aggregate them by
		\begin{equation*}
			S_{\widetilde{\phi}^{'}, null(l)} = \left(S_{\widetilde{\phi}^{'}, null(l)}^{(\widetilde{I})}+S_{\widetilde{\phi}^{'}, null(l)}^{(\widetilde{J})}\right)/2,\quad \widehat{\sigma}_l^2 =  \left(\widehat{\sigma}^2_{(\widetilde{I}),l}+\widehat{\sigma}^2_{(\widetilde{J}),l}\right)/2.
		\end{equation*}
		Calculate the p-value by $2\left(1-\Phi(n^{-1/2}|S_{\widetilde{\phi}^{'},null(l)}|/\widehat{\sigma}_l)\right)$, where $\Phi(\cdot)$ is the cumulative distribution function of a standard normal distribution. 
	\end{algorithm}
	
\section{Theoretical properties}
\label{sec:theory}

In this section, we investigate the asymptotic properties of the proposed procedures under an ERM with non-differentiable loss, potentially involving nuisance parameters. We focus on the uniform validity of the proposed procedures to test a low-dimensional hypothesis. For testing a hypothesis with a growing dimension, we propose a bootstrap procedure and prove its validity in the supplementary material. In the supplementary material, we also provide the convergence rates of $\widehat{\bm \beta}_{\phi}$ and $\widehat{\bm w}_{\phi, l}$.

\subsection{Asymptotic properties without nuisance parameters}
\label{sec:thoery_wo_nusiance}

First, we consider the situation without nuisance parameters.  We assume the following conditions hold on each split dataset in the sample-splitting (and cross-fitting) procedure. For notation simplicity, we omit the subscript indicating the split dataset being used.

\begin{enumerate}[label=(\alph*)]
  \item\label{cond:1} The design matrix is bounded, i.e., $\|\bm X\|_{\infty}\leq M$ with probability $1$; there is a constant $C$ such that $|\bm x^\top\bm\beta^*_{\phi}|\leq C$, and a constant $c>0$ such that $|\beta_{\phi, j_0}^*|\geq c$ for some index $j_0$. Let $f_{x_{j_0}|\bm x_{-j_0}}(x_{j_0} ,a)$ be the conditional density function of $X_{j_0}$ given $\bm X_{-j_0}$ and $A$. We assume that $f_{x_{j_0}|\bm x_{-j_0} ,a}^{\prime}(x_{j_0})$ and $f_{x_{j_0}|\bm x_{-j_0} ,a}^{\prime\prime}(x_{j_0})$ are bounded for both $a=1$ and $-1$. 
  
	\item\label{cond:3} There exists a positive constant $\gamma$ such that for all $t_0>t>0$,
	\begin{equation*}
		\sup_{j, a\in\{-1, 1\}} \mathbb{P}(|t_j-a\bm X^\top\bm\beta^*_{\phi}|\leq t)\leq \tau t^{\gamma},
	\end{equation*}
	where $\tau$ and $t_0$ are some constants.
	
	\item\label{cond:4} We assume that the eigenvalues of
\begin{eqnarray*}
     \bE\left[\left(\Delta_0+\sum_{j=1}^J \Delta_j1\left\{A\bm X^\top\bm\beta^*_{\phi}-t_j\geq 0\right\}\right)^2\bm X\bm X^\top\right]
\end{eqnarray*}
is bounded away from $+\infty$ and $0$ by some constants.
\end{enumerate}

Condition~\ref{cond:1} assumes a bounded design which is a common condition in high-dimensional literature \citep{ning2017, vandegeer2014, chunhui2017}. In addition, Condition~\ref{cond:1} also assumes $\bm\beta_{\phi}^*\not =0$ and some regularity conditions on the conditional density function of the covariates; these conditions are firstly introduced in \citet{koo2008bahadur} and then adopted in \citet{peng2016error, wang2019distributed} to ensure that the hessian of the $L_{\phi}(\bm\beta)$ is well defined and continuous in $\bm\beta$. Condition~\ref{cond:3} assumes that  samples do not concentrate on the jump discontinuity points. This is satisfied when at least one covariate with a non-zero coefficient is continuous and has a bounded density function. Condition~\ref{cond:3} ensures the L-2 convergence of the $1\{t_j-a\bm X^\top\widehat{\bm\beta}_{\phi}\geq 0\}$ to $1\{t_j-a\bm X^\top{\bm\beta}^*_{\phi}\geq 0\}$. Condition~\ref{cond:4} guarantees the uniform convergence of the variance estimator $\widehat{\sigma}_l$.

Now, we provide the uniform validity of the kernel-smoothed decorrelated score under the null hypothesis. 
\begin{theorem}\label{thm:score_null_main}
Denote $s^\prime=\max_l\|\bm w^*_{\phi,l}\|_0$. Assume that $\|\widehat{\bm\beta}_{\phi}-\bm\beta_{\phi}^*\|_2\leq \Delta_{\beta,2}$ with probability approaching to $1$, $s^\prime \sqrt{\log p/(nh_{\rm gb})}=o(1)$ and $\max_l|\bm x_{-l}^\top\bm w^*_{\phi,l}|$ is bounded by $R=o(n^{1/6})$. Taking $\mu_n\asymp\delta_n+Rh_{\rm gb}^2+R\Delta_{\beta,2}$, where $\delta_n=R(\log p/(nh_{\rm gb}))^{1/2}$.
	Further assume that $\sqrt{n}R(h_{\rm lo}+\sqrt{\log p}\Delta_{\beta,2}^{2\gamma/(\gamma+2)})=o(1)$, $\sqrt{n}(s^\prime \mu_n)(\sqrt{\log p/n}+h_{\rm lo}+\Delta_{\beta, 2})=o(1)$, and $(Rs'\mu_n+R^2\sqrt{\log p/n}+R^2\Delta_{\beta,2}+R^2h_{\rm lo})\sqrt{\log p}=o(1)$. If Conditions~\ref{cond:1} -~\ref{cond:4} are satisfied,  under the null hypothesis, we have
	\begin{eqnarray*}
			&&\max_{l\in \mathcal{H}_0}\sup_{\alpha\in (0,1)}\left|\mathbb{P}\left(\left|n^{1/2}\widehat{\sigma}_l^{-1}S_{\widetilde{\phi}^{'}, null(l)}\right|\leq \Phi^{-1}(1-\alpha/2)\right)-(1-\alpha)\right|=o_p(1),
\end{eqnarray*}
	where $\mathcal{H}_0$ is the index set of zero coefficients in $\bm\beta_{\phi}^*$ under the null hypothesis.
\end{theorem}

Define
\begin{eqnarray*}
	&&{\sigma}_l^2 =\bE\left[\left(\Delta_0+\sum_{j=1}^J \Delta_j1\left\{A\bm X^\top\bm\beta^*_{\phi}-t_j\geq 0\right\}\right)^2(X_l-\bm X_{-l}^\top \bm w^*_{\phi,l})^2\right].
\end{eqnarray*} 
Theorem~\ref{thm:score_null_main} implies that the asymptotic variance ${\sigma}_l^2$ can be estimated by $\widehat{\sigma}_l^2$ in Algorithm~\ref{algorithm:1}. 

The following corollary provides the uniform validity of the confidence interval constructed through the one-step debiased estimator $\widetilde{\beta}_{\phi, l}$.
\begin{corollary}\label{thm:coverage_main}
	Assume the same conditions in Theorem~\ref{thm:score_null_main}, we have
	\begin{eqnarray*}
			&&\max_{l}\sup_{\alpha\in (0,1)}\left|\mathbb{P}\left(\left|n^{1/2}\widehat{\sigma}_l^{-1}\widehat{I}_l\left(\widetilde{\beta}_{\phi, l}-{\beta}^*_{\phi, l}\right)\right|\leq \Phi^{-1}(1-\alpha/2)\right)-(1-\alpha)\right|=o_p(1).
\end{eqnarray*}
\end{corollary}
Define
\begin{eqnarray*}
	\widetilde{\sigma}_l^2 &=& {\sigma}_l^2/I_l^2,\quad
	I_l^2 = \bE\left[\sum_{j} \Delta_j \delta(t_j-A\bm X^\top\bm\beta^*_{\phi})(X_l-\bm X_{-l}^\top\bm w^*_{\phi,l})^2\right].
\end{eqnarray*}
Corollary~\ref{thm:coverage_main} implies that the asymptotic variance $\widetilde{\sigma}_l^2$ can be estimated by $\widehat{\sigma}_l^2/\widehat{I}_l^2$.

Compared with the theoretical conditions for the marginal validity of the decorrelated score under a differentiable strictly convex loss function without nuisance parameters, our conditions for the uniform validity under a non-differentiable loss function assumes a more sparse model. To see this, \citet{ning2017} show that the condition required for the marginal validity of the decorrelated score is $\max\{s^\prime, s^*\}\log p/\sqrt{n}\to 0$; this is equivalent to $\sqrt{n}\|\widehat{\bm w}_{\phi,l}-\bm w^*_{\phi,l}\|_2^2\to 0$ and $\sqrt{n}\Delta_{\beta,2}^2\to 0$. For the hinge loss, Theorem~\ref{thm:score_null_main} requires that $\sqrt{n}s^\prime \mu_n\Delta_{\beta, 2}\to 0$ in addition to $\sqrt{n}\Delta_{\beta,2}^2\to 0$; this is equivalent to $\sqrt{n}\|\widehat{\bm w}_{\phi,l}-\bm w^*_{\phi,l}\|_1\Delta_{\beta, 2}\to 0$ and $\sqrt{n}\Delta_{\beta,2}^2\to 0$ (see Appendix for the convergence rate of $\widehat{\bm w}$). If $\|\widehat{\bm w}_{\phi,l}-\bm w^*_{\phi,l}\|_2^2\lesssim \|\widehat{\bm w}_{\phi,l}-\bm w^*_{\phi,l}\|_1\Delta_{\beta, 2}$, the conditions in Theorem~\ref{thm:score_null_main} and~\ref{thm:score_null_nuisance} are more restrictive than \citet{ning2017}. In Appendix (the Proof of Theorem~\ref{thm:score_null_nuisance}), we show that under a dedicated sample-splitting algorithm, we can reduce this requirement to a weaker condition than \citet{ning2017}. With this sample-splitting algorithm, we only require that $\sqrt{n}\|\widehat{\bm w}_{\phi,l}-\bm w^*_{\phi,l}\|_2\Delta_{\beta,2}\to 0$ in addition to $\sqrt{n}\Delta_{\beta,2}^2\to 0$. The requirements that $\sqrt{n}\|\widehat{\bm w}_{\phi,l}-\bm w^*_{\phi,l}\|_2\Delta_{\beta,2}\to 0$ and  $\sqrt{n}\Delta_{\beta,2}^2\to 0$ are weaker than $\sqrt{n}\|\widehat{\bm w}_{\phi,l}-\bm w^*_{\phi,l}\|_2^2\to 0$ and $\sqrt{n}\Delta_{\beta,2}^2\to 0$, when $\Delta_{\beta,2}\lesssim \|\widehat{\bm w}_{\phi,l}-\bm w^*_{\phi,l}\|_2$. However, the dedicated sample-splitting algorithm leads to a significant increase in computation time. Hence, we mainly focus on the proposed procedure in the paper.

\subsection{Asymptotic properties with nuisance parameters}
\label{sec:thoery_w_nusiance}

In this section, we investigate the theoretical property of the proposed inference procedure when nuisance parameters exist. 
\begin{enumerate}[label=(\alph*)]
\addtocounter{enumi}{3}
	\item\label{cond:5} There is a constant $C$ such that $\max\left\{W_1, W_{-1}\right\}\leq C$. $\partial_{x_{j_0}}W_a(\bm x)$ and $\partial^2_{x_{j_0}}W_a(\bm x)$ are bounded. 	
	\item\label{cond:2} There exist positive constants $\eta$ and $\zeta$ such that 
	\begin{eqnarray*}
		\sup_{\bm x, a, y}\left|\widehat{W}_a-\widebar{W}_a\right|&=&O_p(n^{-\zeta}), \\
  \sup_{\bm x}\left|\bE\left[\widehat{W}_a\mid \bm X=\bm x\right]-W_a\right|&=&O_p(n^{-\eta}),
	\end{eqnarray*}
	where $\widebar{W}_a$ is the point-wise limit of $\widehat{W}_a$ as $n\to \infty$.
\end{enumerate}

Condition~\ref{cond:5} assumes that the weights are bounded and smooth. Condition~\ref{cond:2} assumes that the convergence rates of the nuisance parameters are upper bounded by $O_p(n^{-\zeta})$; the expectation of the weights with estimated nuisance parameters approximates $W_a(\bm X)$ faster than $O_p(n^{-\eta})$.

We now provide the uniform validity of the kernel-smoothed decorrelated score under the null hypothesis when there exist nuisance parameters. 
\begin{theorem}\label{thm:score_null_nuisance}
	Denote $s^\prime=\max_l\|\bm w^*_{\phi,l}\|_0$ and $\delta_n=R(\log p/(nh_{\rm gb}))^{1/2}+Rn^{-\eta}/h_{\rm gb}$. Assume that $\|\widehat{\bm\beta}_{\phi}-\bm\beta_{\phi}^*\|_2\leq \Delta_{\beta,2}$ with probability approaching to $1$, $s^\prime \sqrt{\log p/(nh_{\rm gb})}=o(1)$ and $\max_l|\bm x_{-1}^\top\bm w^*_{\phi,l}|$ is bounded by $R=o(n^{1/6})$. Taking $\mu_n\asymp\delta_n+Rh_{\rm gb}^2+R\Delta_{\beta,2}$. Further assume that $\sqrt{n}R(h_{\rm lo}+n^{-\eta}+\sqrt{\log p}\Delta_{\beta,2}^{2\gamma/(\gamma+2)})=o(1)$, $\sqrt{n}(s^\prime \mu_n)(n^{-\eta}+\sqrt{\log p/n}+h_{\rm lo}+\Delta_{\beta, 2})=o(1)$, and $(Rs'\mu_n+R^2\sqrt{\log p/n}+R^2n^{-\zeta}+R^2\Delta_{\beta,2}+R^2h_{\rm lo})\sqrt{\log p}=o(1)$. If Conditions~\ref{cond:1} -~\ref{cond:2},  under the null hypothesis, we have
	\begin{eqnarray*}
			&&\max_{l\in \mathcal{H}_0}\sup_{\alpha\in (0,1)}\left|\mathbb{P}\left(\left|n^{1/2}\widehat{\sigma}_l^{-1}S_{\widetilde{\phi}^{'}, null(l)}\right|\leq \Phi^{-1}(1-\alpha/2)\right)-(1-\alpha)\right|=o_p(1),
\end{eqnarray*}
	where $\mathcal{H}_0$ is the index set of zero coefficients in $\bm \beta_{\phi}^*$ under the null hypothesis and
\begin{eqnarray*}
	{\sigma}_l^2 &=& \bE\left[\left\{\sum_{a}aW_a\left(\Delta_0+\sum_{j=1}^J \Delta_j1\left\{a\bm X^\top\bm\beta^*_{\phi}-t_j\geq 0\right\}\right)\right\}^2(X_l-\bm X_{-l}^\top \bm w^*_{\phi,l})^2\right].
\end{eqnarray*} 
\end{theorem}
Compared with Theorem~\ref{thm:score_null_main}, the $\delta_n$ also involves the $Rn^{-\eta}/h_{\rm gb}$ due to the additional nuisance parameters.

\begin{corollary}\label{thm:coverage_nuisance}
	Assume the same conditions in Theorem~\ref{thm:score_null_nuisance}, we have
	\begin{eqnarray*}
			&&\max_{l}\sup_{\alpha\in (0,1)}\left|\mathbb{P}\left(\left|n^{1/2}\widehat{\sigma}_l^{-1}\widehat{I}_l\left(\widetilde{\beta}_{\phi, l}-{\beta}^*_{\phi, l}\right)\right|\leq \Phi^{-1}(1-\alpha/2)\right)-(1-\alpha)\right|=o_p(1).
\end{eqnarray*}
\end{corollary}

Define
\begin{eqnarray*}
	\widetilde{\sigma}_l^2 &=& {\sigma}_l^2/I_l^2,\quad
	I_l^2 = \bE\left[\sum_{a}\sum_{j} \Delta_j W_a\delta(t_j-a\bm X^\top\bm\beta^*_{\phi})(X_l-\bm X_{-l}^\top\bm w^*_{\phi,l})^2\right].
\end{eqnarray*}
Theorem~\ref{thm:score_null_nuisance} and Corollary~\ref{thm:coverage_nuisance} imply that the asymptotic variance component $\sigma_l^2$ and $\sigma_l^2/I_l^2$ can be estimated by $\widehat{\sigma}_l^2$ and $\widehat{\sigma}_l^2/\widehat{I}_l^2$, respectively. They require that $R^2n^{-\zeta}\sqrt{\log p}=o(1)$, where $n^{-\zeta}$ corresponds to the slowest convergence rate of the nuisance parameters. This ensures that the $\widehat{I}_l$ converges fast enough to construct a uniformly valid testing procedure/debiased estimator. Furthermore, compared with Theorem~\ref{thm:score_null_main}, Theorem~\ref{thm:score_null_nuisance} also requires that $\sqrt{n}Rn^{-\eta}\to 0$ and $\sqrt{n}(s^\prime \mu_n)n^{-\eta}\to 0$. These requirements ensure that the uncertainty of estimating the nuisance parameters is asymptotically ignorable, and does not affect the asymptotic variance. Under a doubly robust formulation, these requirements are not restrictive and can be satisfied even when the nuisance parameter estimates have a slow convergence rate (see below for concrete examples). 

In Theorem~\ref{thm:score_null_nuisance} and Corollary~\ref{thm:coverage_nuisance}, we require that $\sqrt{n}Rn^{-\eta}\to 0$. 
In classification with missing labels example, assume that there exists $\alpha,\beta>0$ such that $\sup_{\bm x}|\widehat{\pi}_1(\bm x)- \widebar{\pi}(\bm x)|=O_p(n^{-\alpha})$ and $\sup_{\bm x}|\widehat{p}_a(\bm x)- \widebar{p}_a(\bm x)|=O_p(n^{-\beta})$. When we assume that both $\widebar{\pi}_1=\pi_1$ and $\widebar{p}_a=p_a$, the condition that $\sqrt{n}Rn^{-\eta}\to 0$ holds with $\eta=\alpha+\beta$ if $\sqrt{n}Rn^{-\alpha-\beta}\to 0$, since 
\begin{eqnarray*}
    \sup_{\bm x}\left|\bE[\widehat{W}_a(\bm X, A;\widehat{\pi}, \widehat{p})\mid \bm X=\bm x]-W_a(\bm x)\right|=O_p(n^{-\alpha-\beta}).
\end{eqnarray*}
When the missing mechanism is known, i.e., $\widehat{\pi}_1=\pi_1$ is known, the condition that $\sqrt{n}Rn^{-\eta}\to 0$  automatically holds since
\begin{eqnarray*}
    \sup_{\bm x}\left|\bE[\widehat{W}_a(\bm X, A;\widehat{\pi}, \widehat{p})\mid \bm X=\bm x]-W_a(\bm x)\right|=0,
\end{eqnarray*}
for any $\widehat{p}_a$.

In the inference of individualized treatment rule example, assume that there exists $\alpha,\beta>0$ such that $\sup_{\bm x}|\widehat{p}_a(\bm x)- \widebar{p}_a(\bm x)|=O_p(n^{-\alpha})$ and $\sup_{\bm x}|\widehat{Q}_a(\bm x)- \widebar{Q}_a(\bm x)|=O_p(n^{-\beta})$. When we assume that both $\widebar{p}_a=p_a$ and $\widebar{Q}_a=Q_a$, the condition $\sqrt{n}Rn^{-\eta}\to 0$ holds with $\eta=\alpha+\beta$ if $\sqrt{n}Rn^{-\alpha-\beta}\to 0$ since
\begin{eqnarray*}
    \sup_{\bm x}\left|\bE[\widehat{W}_a(\bm X, A;\widehat{p}, \widehat{Q})\mid \bm X=\bm x]-W_a(\bm x)\right|=O_p(n^{-\alpha-\beta}).
\end{eqnarray*}
When the treatment assignment mechanism is known, i.e., $\widehat{p}_a=p_a$ is known (for example, in randomized clinical trials), the condition that $\sqrt{n}Rn^{-\eta}\to 0$  automatically holds since
\begin{eqnarray*}
    \sup_{\bm x}\left|\bE[\widehat{W}_a(\bm X, A;\widehat{p}, \widehat{Q})\mid \bm X=\bm x]-W_a(\bm x)\right|=0,
\end{eqnarray*}
for any $\widehat{Q}_a$.

With a bounded $R$, requirement on $\sqrt{n}Rn^{-\alpha-\beta}\to 0$ is equivalent to $\alpha+\beta>1/2$, which can be satisfied by many estimation methods. For example, in the inference of the individualized treatment rule example, if we adopt generalized linear models with lasso penalties to estimate the propensity score and the outcome regression models, it is satisfied when $\widetilde{s}\log p/\sqrt{n}=o(1)$, where $\widetilde{s}$ is an upper bound of the number of the non-zero coefficients in the propensity score and the outcome regression models. If the propensity score is estimated by a regression spline estimator and is known to be $p_{\pi}$-dimensional (low-dimensional) by design, we have $\alpha=1/3$ if $\pi$ belongs to the H\"older class with a smoothness parameter greater than $5p_{\pi}$ \citep{newey1997}. In this case, we only need $\beta>1/6$.

\section{Simulation}
\label{sec:sim}

In this section, we conduct simulation studies to examine the performance of the proposed inference procedure. We consider two simulation scenarios: classification without nuisance parameters (see Section~\ref{sec:common_class}) and estimating ITR with nuisance parameters (see Section~\ref{sec:class}). Specifically, the population is a mixture of two subgroups with a probability $0.4$ from Group I and probability $0.6$ from Group II. For Group I, the covariate vector is generated from $N(\xi\bm \mu_0, I_{p\times p}-0.1\bm e_1\bm e_1^\top)$; for Group II, the covariate vector is generated from $N(\xi\bm \mu_1, I_{p\times p})$, where $\mu_0=(-1,1,-0.5,0.5,0,\cdots,0)^\top$ and $\mu_1=(1,-1,-1,-1,0,\cdots, 0)^\top$.
\begin{enumerate}[label=\Roman*.]
	\item \label{sim:1}  The label for Group I is $A=1$, and the label for Group II is $A=-1$. The goal is to classify Group I from Group II based on the data;
	\item \label{sim:2} The treatment assignment mechanism follows $P(A=1\mid \bm X)=\exp(0.25\times (X_1^2+X_2^2+X_1X_2)/(1+0.25\times (X_1^2+X_2^2+X_1X_2))$. The observed outcome $Y=Y(a)$ if the treatment $a$ is assigned to the patient, $Y(a)= (\bm X^\top\bm\gamma)^2 + C(\bm X; G) I(a=1) +\epsilon$, where $\gamma=(-0.4,-0.4,0.4,-0.4,0,\cdots, 0)^\top$ and $\epsilon$ follows a standard normal distribution. Here, $C(\bm X;G=1)  =|X_1|+0.5$ for Group I patients and $C(\bm X;G=2) = -(|X_1|+0.5)$ for Group II patients. The goal is to estimate the optimal individualized treatment rules that maximize the outcome. The CATE is given by $\mathbb{E}[C(\bm X; G)\mid \bm X]$; and thus, by calculation, the sign of CATE is the same as $\mathbb{P}(G=1\mid \bm X)-0.5$.
\end{enumerate} 
Our Scenario II assumes that there are two subgroups with different treatment effects, and the group membership $G$ is unobserved. Similar setups have been adopted in recent literature for heterogeneous subgroup-level treatment effects \citep{lin2021bags, chandra2023bayesian}. For each scenario, the parameter $\xi$ controls the magnitude of the overlapping of two subgroups. Two subgroups are easier to separate by a linear decision rule for a larger $\xi$. We gradually increase $\xi$ from $0.1$ to $1$ (by 0.1) and the dimension of the covariate $p$ from $500$ to $1600$, which lead to $30$ settings in total for each scenario. 

We compare our proposed method with an ad-hoc method and the method in \cite{liang2022estimation} using a logistic loss. For the ad-hoc approach, we first use the hinge loss with the lasso penalty to identify covariates with non-zero coefficient estimates. Then, we refit the hinge loss without any penalty using the identified covariates and first 8 covariates, and employ the inference procedure for low-dimensional settings \citep{koo2008bahadur} to construct test statistics and confidence intervals for identified and first 8 covariates. In Scenario II, we further combine the ad-hoc method with the cross-fitting algorithm proposed in \citet{chernozhukov2017double} as the competitor.  For Scenario~II, we use the kernel regression after the variable screening to estimate nuisance parameters for both the proposed and ad-hoc methods. For each simulation setting,  we repeatedly simulate the training samples $500$ times, the sample sizes of which vary from $500$ to $1600$. To evaluate the inference procedure, we report type I error (testing $\mathcal{H}_0:\beta_{\phi, l}^*=0$, where $l=5,6,7,8$), power (testing $\mathcal{H}_0:\beta_{\phi, l}^*=0$, where $l=1,2,3,4$), the averaged coverage of the $95\%$ confidence intervals for the eight coordinates in $\bm\beta^*_{\phi}$, and the averaged length of the $95\%$ confidence intervals. The true value of $\bm\beta^*_{\phi}$ is determined by the average over $500$ replicates with $n=2500$. The coordinates of $\bm\beta^*_{\phi}$ are set to zero if the absolute value of its average estimates is less than 0.01. When comparing interval length with the method using a logistic loss, we normalize the interval length based on the first coordinate of $\bm\beta^*_{\phi}$. In addition, we also report the classification accuracy (for Scenario I), value function (for Scenario II), and estimation errors of debiased one-step estimator (against $\bm\beta^*_{\phi}$ under hinge loss) to compare the methods using a hinge loss vs. a logistic loss.

\begin{figure}
\centering
{\includegraphics[width=\linewidth]{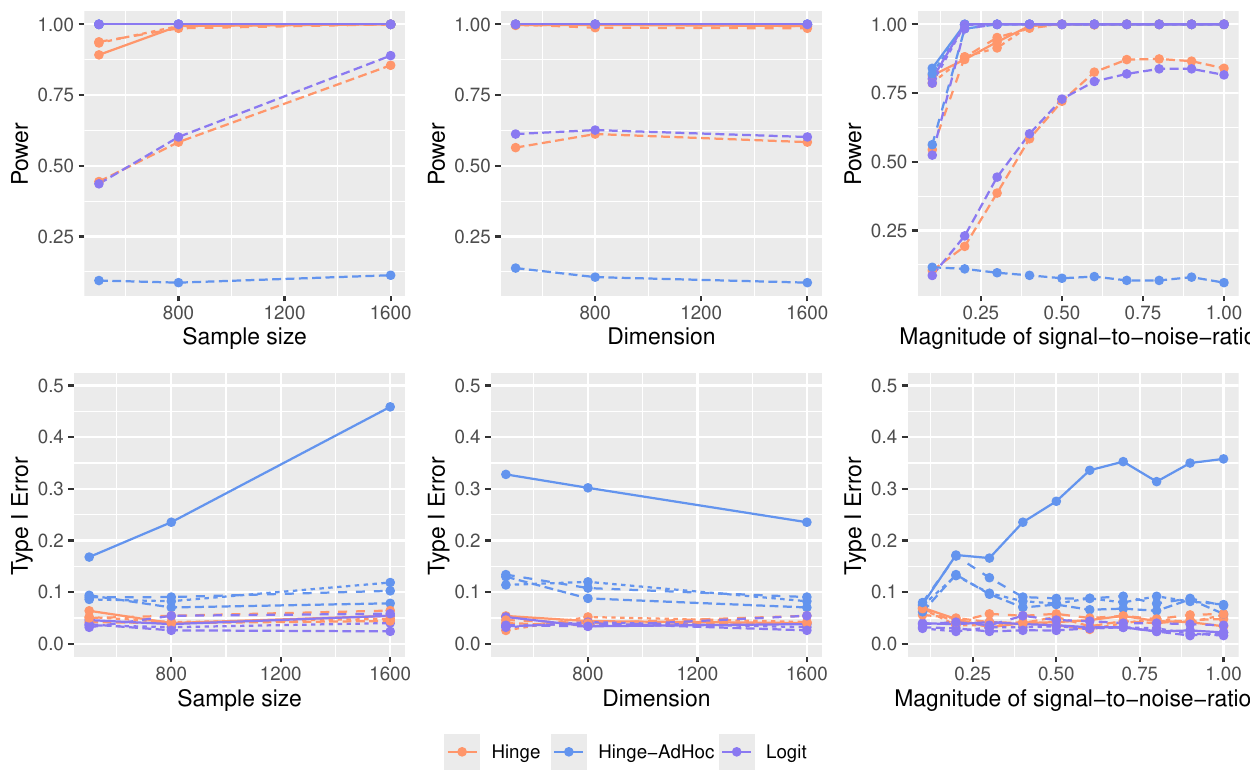}}
\caption{Testing results for Scenario~I with the change of sample size when $\xi=0.4$ and $p=800$, the change of $\xi$ ($n=800$; $p=1600$) and the change of $p$ ($n=800$; $\xi=0.4$). Line styles represent different coefficients.\\}\label{fig:testing1}
\end{figure}

\begin{figure}
\centering
{\includegraphics[width=\linewidth]{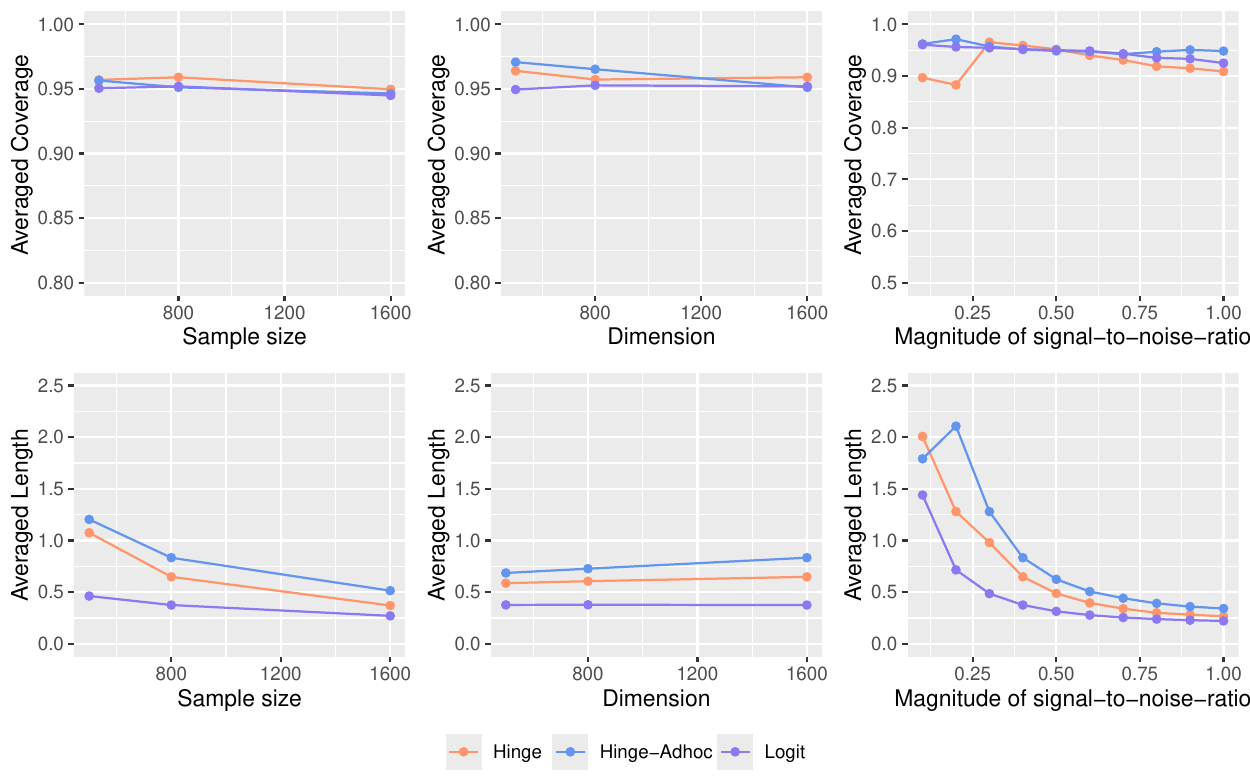}}
\caption{Coverage results for Scenario~I with the change of sample size when $\xi=0.4$ and $p=800$, the change of $\xi$ ($n=800$; $p=1600$) and the change of $p$ ($n=800$; $\xi=0.4$).\\}\label{fig:coverage1}
\end{figure}

\begin{figure}
\centering
{\includegraphics[width=\linewidth]{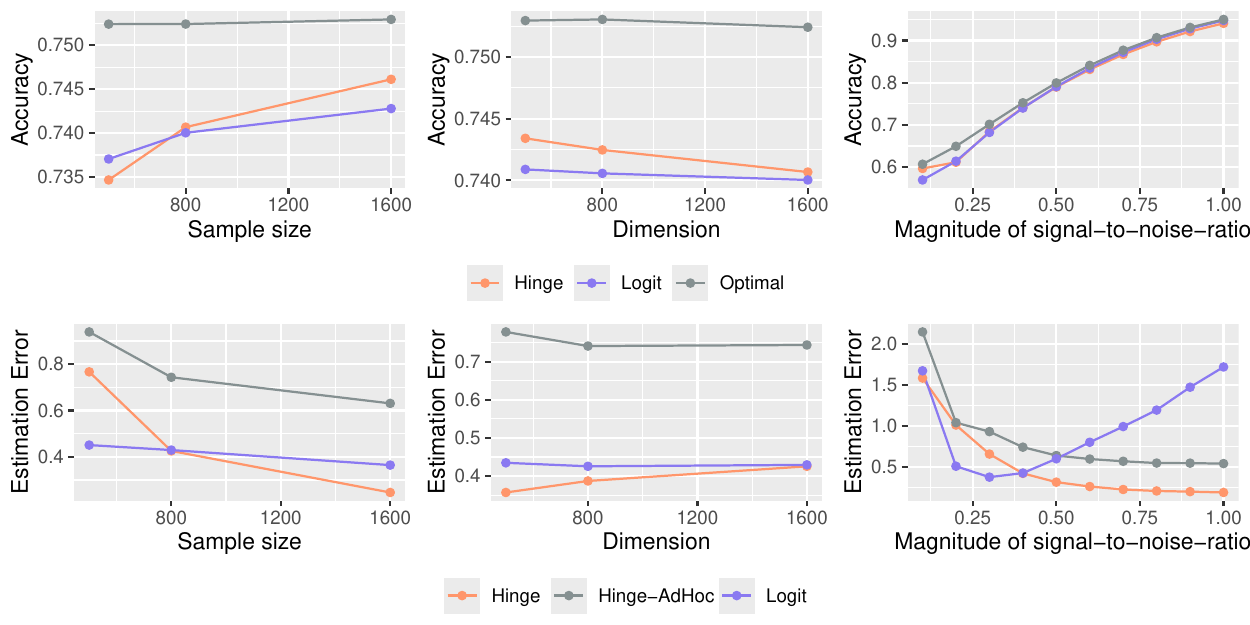}}
\caption{Classification accuracy and estimation error for Scenario~I with the change of sample size when $\xi=0.4$ and $p=800$, the change of $\xi$ ($n=800$; $p=1600$) and the change of $p$ ($n=800$; $\xi=0.4$).\\}\label{fig:estimate1}
\end{figure}

\begin{figure}
\centering
{\includegraphics[width=\linewidth]{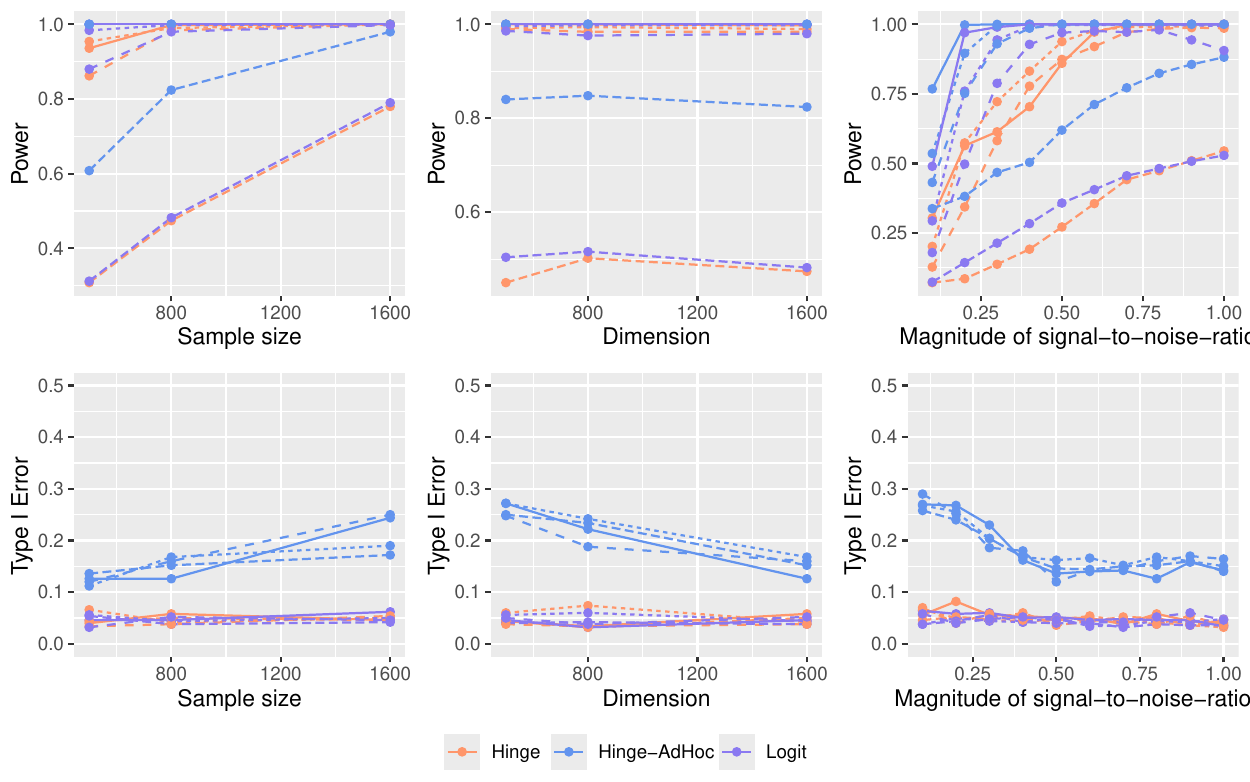}}
\caption{Testing results for Scenario~II with the change of sample size when $\xi=0.8$ and $p=800$, the change of $\xi$ ($n=800$; $p=1600$) and the change of $p$ ($n=800$; $\xi=0.8$). Line styles represent different coefficients.\\}\label{fig:testing2}
\end{figure}

\begin{figure}
\centering
{\includegraphics[width=\linewidth]{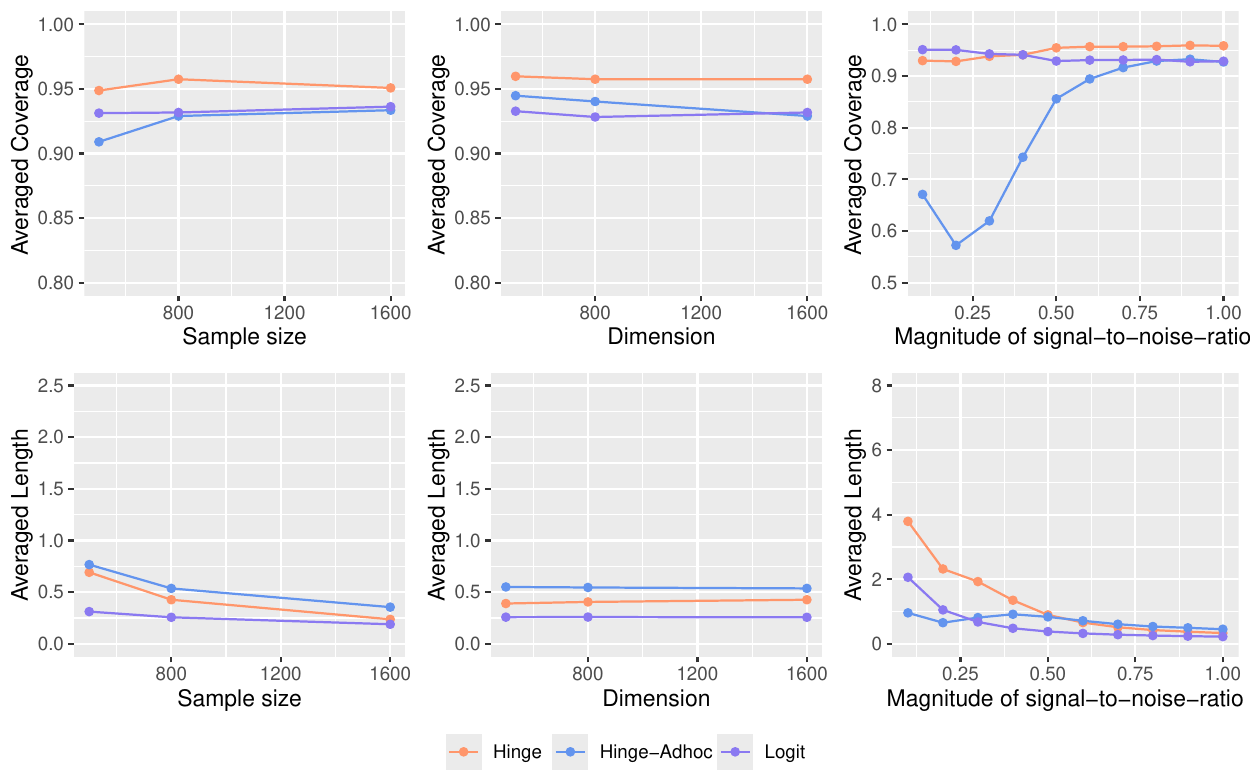}}
\caption{Coverage results for Scenario~II with the change of sample size when $\xi=0.8$ and $p=800$, the change of $\xi$ ($n=800$; $p=1600$) and the change of $p$ ($n=800$; $\xi=0.8$).}\label{fig:coverage2}
\end{figure}

\begin{figure}
\centering
{\includegraphics[width=\linewidth]{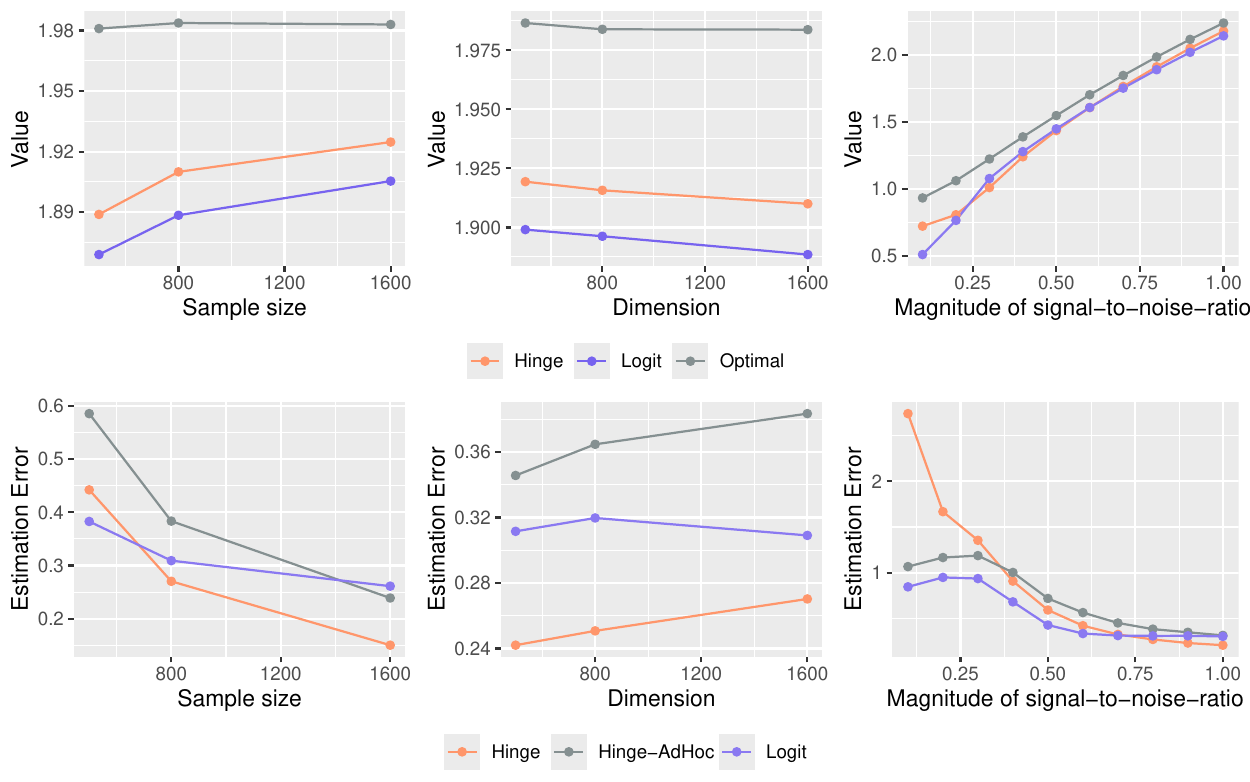}}
\caption{Value function and estimation error for Scenario~II with the change of sample size when $\xi=0.8$ and $p=800$, the change of $\xi$ ($n=800$; $p=1600$) and the change of $p$ ($n=800$; $\xi=0.8$).\\}\label{fig:estimate2}
\end{figure}

Figures~\ref{fig:testing1} - \ref{fig:estimate1} show the simulation results for Scenario~I. The results show that the proposed method has controlled Type-I errors and higher powers than the ad-hoc method. The ad-hoc method has inflated type-I errors, which indicate that the decorrelation and sample-splitting procedures to construct valid test statistics are necessary. From Figure~\ref{fig:coverage1}, the proposed method achieves nominal coverages and has shorter confidence intervals than the ad-hoc method across all settings. When comparing with the method using a logistic loss, our method yields slightly lower power but comparable or higher accuracy, as shown in Figure~\ref{fig:estimate1}. We also observe that the averaged coverage shows a non-monotonic pattern for our proposed method. When $\xi$ changes, the value of $\bm\beta^*_{\phi}$ may also change. The varying $\bm\beta^*_{\phi}$' may contribute to this non-monotonic pattern. The simulation results for Scenario~II are presented in Figures~\ref{fig:testing2} -~\ref{fig:estimate2}. Again, the proposed method yields better results in terms of Type I error control, coverage, and the length of confidence intervals. The proposed method also achieves a higher value compared with the method using a logistic loss. Results of more simulation settings can be found in Appendix D.


\section{Real data examples}
\label{sec:real_data}

 In this section, we apply our proposed inference approach in real world problems. Specifically, we consider two scientific questions: 1) whether we can identify the risk factors associated with uncontrolled HbA1c in a year, given the patients baseline characteristics; 2) whether we can identify driving factors to inform better treatment strategies. 
 
The data we used comes from the electronic medical records linked with Medicare claims data on Type-II diabetic patients with complex commodities. These data are collected through the Heath Innovation Program at the University of Wisconsin. It contains $n=9101$ patients with many covariate information. In order to answer the research questions highlighted above, we include $40$ covariates including sociodemographic variables, disease history, baseline HbA1c levels, etc., and their first-order interactions in our analysis. As a variable screening procedure, we rank these covariates and their interactions by the variances and select $p=120$ covariates with highest variances.The outcome of interest in both questions is whether the patient successfully controlled his or her HbA1c below $8\%$ after one year. 

\subsection{Identify risk factors associated with uncontrolled HbA1c for Type-II diabetic patients}
\label{subsec:real_data_clas}	

Our goal is to identify patients and risk factors associated with uncontrolled HbA1c after one-year of follow-up if following current clinical guideline. This can be considered as a classification problem. The label to predict, denoted as $A$, indicates whether patients have uncontrolled HbA1c after one-year of follow-up. Specifically, we set $A=1$ if patients successfully control the HbA1c level after one-year follow-up; and set $A=-1$, otherwise.

We first use the linear SVM with a lasso penalty to estimate the decision rule and then conduct inference on the estimated decision rule. Under a cross-validation procedure, the estimated decision rule achieves a prediction accuracy of $0.895$ with a standard deviation of $0.003$. After controlling the false discovery rate (FDR) at 0.05 by Benjamini–Hochberg procedure \citep{benjamini1995controlling}, we find that patients with valvular disease are more likely to suffer from uncontrolled HbA1c's (see Table~\ref{tab:coef_realdata_class}); minorities with hypertension are more likely to have HbA1c under control after treatment, which is probably due to the full consideration for patients with hypertension as one common comorbidity in current clinical guideline. After controlling for the FDR, the ad-hoc method reveals that 112 covariates out of 120 covariates are significant. However, when evaluating the length of the constructed confidence intervals, the average interval length for the proposed method is much shorter at $0.122$ compared to that of the ad-hoc method, which is $0.244$. The high number of identified risk factors and long length of confidence intervals may be attributed to biased test statistics and interval estimations, as discussed in Section~\ref{sec:sim}.

\begin{table}
	\caption{Coefficients and p-value for the identified significant covariates of the estimated decision rule after FDR control.}
	\label{tab:coef_realdata_class}
	{\footnotesize \begin{center}
		\begin{tabular}{lccc}
			\toprule
			\multicolumn{1}{c}{Covariate} & Coef &  P-value &95\% - CI\\
			\hline 
Valvular disease (Yes) & -2.079 & $1.528\times 10^{-99}$ & [-2.118, -2.041]\\
			Hypertension (Yes) : Race (other) & 2.126 & $3.687\times 10^{-88}$ & [2.091, 2.161]\\			\bottomrule
		\end{tabular}
\end{center}}
\end{table}

\subsection{Identify driving factors of the estimated ITR to inform better clinical guideline for Type-II diabetic patients}
\label{subsec:real_data_itr}

We aim to estimate the optimal ITR and identify the driving factors  to inform future clinical guideline for Type-II diabetic Medicare patients. We consider two treatment options: hypoglycemic agents versus usual care. We set $A=1$ if the patient received hypoglycemic agent at baseline and $A=-1$ otherwise. The outcome $Y$ is whether the patient successfully controlled his or her HbA1c below $8\%$ after one year. Under these specifications, the weighted classification problem proposed in Section~\ref{sec:class} aims to estimate an ITR such that under the derived ITR, the chance of successfully controlling HbA1c below $8\%$ after one year is maximized for each individual.

 To demonstrate that the estimated ITR informs a better treatment strategy than the current clinical practice, we use the cross-validation procedure to calculate the chance of successfully controlling HbA1c if the estimated ITR were implemented. The results show that the current clinical practice has a success rate of 0.860 with a standard deviation of 0.008; the estimated ITR has a success rate of 0.871 with a standard deviation of 0.013. In terms of the identified driving factors of the estimated ITR, using the proposed inference method, after controlling FDR at 0.05, we find that patients having hypertension, especially with higher A1c levels at baseline, are more likely to benefit from hypoglycemic agents in controlling HbA1c, which confirmed the importance of hypertension in Type-II diabetes care. In addition, we also find that female patients with chronic complications, are more likely to benefit from hypoglycemic agents in controlling HbA1c, see Table~\ref{tab:coef_realdata}. We also implement the ad-hoc method as a comparison. After controlling for the FDR, the ad-hoc method identifies 107 significant driving factors out of 120 predictors. The averaged length of confidence intervals for the ad-hoc method is 0.423, longer than that of the proposed method, which is 0.122. 

\begin{table}
	\caption{Coefficients and p-value for the identified significant covariates of the estimated optimal ITR after FDR control. Special chronic conditions refer to chronic conditions including amputation, chronic blood loss, drug abuse, lymphoma, metastatic cancer, and peptic ulcer disease. }
	\label{tab:coef_realdata}
	{\footnotesize \begin{center}
		\begin{tabular}{lccc}
			\toprule
			\multicolumn{1}{c}{Covariate} & Coef &  P-value &95\% - CI\\
			\hline 
			Hypertension (Yes) & -0.0451 & $8.32\times 10^{-4}$ & [-0.0918, 0.0016]\\
			Chronic Complications (Yes) : Female & 0.1244 & $4.51\times 10^{-6}$ & [0.0947, 0.154]\\
			Hypertension (Yes) : Baseline A1c & 0.0627 & $6.95\times 10^{-7}$ & [0.0205, 0.105]\\	
			\bottomrule
		\end{tabular}
\end{center}}
\end{table}

\section{Discussion}
\label{sec:diss}

We propose a high-dimensional inference procedure for a non-differentiable convex loss function in a general classification framework, which can be utilized to discover the driving factors in decision making. 
In particular, combined with the cross-fitting algorithm, our procedure can accommodate weights involving additional nuisance parameters, which may be estimated via nonparametric or other machine learning algorithms. 

There are multiple directions that could be further studied in the future. First, although we allow a non-differentiable surrogate loss, we still require the convexity of the surrogate loss. It would be interesting to see how the proposed method can be extended to deal with a non-convex surrogate loss or even zero-one loss without any relaxation. Second, we may consider distributed inference or online inference. When the sample size is large, the computation may be infeasible on a single machine due to limited resources. In this case, distributed inference can leverage the computation power of a cluster of machines and reduce the runtime. When the sample size is limited, it would be important to understand how to utilize newly collected samples to improve the learning process of the decision rule. Third, in this work,  we only consider bounded designs; we may consider extending these results to sub-gaussian designs or designs with heavy tails.

\newpage
\begin{center}
	{\large\bf Supplemental Materials}
\end{center}

\begin{description}
	
	\item[Supplementary material:] Supplementary material contains the proofs of all theorems, an extension to testing high-dimensional hypothesis, and the convergence rate of $\widehat{\bm\beta}_{\phi}$ when using a hinge loss. (pdf file)
	
	\item[R-package for SVM Inference:] R-package SVMInference containing code to perform the proposed methods (for hinge loss) described in the article. (GNU zipped tar file)		
\end{description}

\bibliographystyle{apalike}
\bibliography{ref.bib}

@article{bing2023optimal,
  title={Optimal discriminant analysis in high-dimensional latent factor models},
  author={Bing, Xin and Wegkamp, Marten},
  journal={Annals of Statistics},
  volume={51},
  number={3},
  pages={1232--1257},
  year={2023},
  publisher={Institute of Mathematical Statistics}
}

@article{dobriban2018high,
  title={High-dimensional asymptotics of prediction: Ridge regression and classification},
  author={Dobriban, Edgar and Wager, Stefan},
  journal={Annals of Statistics},
  volume={46},
  number={1},
  pages={247--279},
  year={2018},
  publisher={JSTOR}
}

@article{fan2008high,
  title={High dimensional classification using features annealed independence rules},
  author={Fan, Jianqing and Fan, Yingying},
  journal={Annals of statistics},
  volume={36},
  number={6},
  pages={2605},
  year={2008}
}

@article{liang2022estimation,
  title={Estimation and inference on high-dimensional individualized treatment rule in observational data using split-and-pooled de-correlated score},
  author={Liang, Muxuan and Choi, Young-Geun and Ning, Yang and Smith, Maureen A and Zhao, Ying-Qi},
  journal={Journal of Machine Learning Research},
  volume={23},
  number={262},
  pages={1--65},
  year={2022}
}

@article{chandra2023bayesian,
  title={Bayesian nonparametric common atoms regression for generating synthetic controls in clinical trials},
  author={Chandra, Noirrit Kiran and Sarkar, Abhra and de Groot, John F and Yuan, Ying and M{\"u}ller, Peter},
  journal={Journal of the American Statistical Association},
  volume={118},
  number={544},
  pages={2301--2314},
  year={2023},
  publisher={Taylor \& Francis}
}

@article{lin2021bags,
  title={BAGS: A Bayesian adaptive group sequential trial design with subgroup-specific survival comparisons},
  author={Lin, Ruitao and Thall, Peter F and Yuan, Ying},
  journal={Journal of the American Statistical Association},
  volume={116},
  number={533},
  pages={322--334},
  year={2021},
  publisher={Taylor \& Francis}
}

@article{cai2025semi,
  title={Semi-supervised triply robust inductive transfer learning},
  author={Cai, Tianxi and Li, Mengyan and Liu, Molei},
  journal={Journal of the American Statistical Association},
  volume={120},
  number={550},
  pages={1037--1047},
  year={2025},
  publisher={Taylor \& Francis}
}

@article{deng2024optimal,
  title={Optimal and safe estimation for high-dimensional semi-supervised learning},
  author={Deng, Siyi and Ning, Yang and Zhao, Jiwei and Zhang, Heping},
  journal={Journal of the American Statistical Association},
  volume={119},
  number={548},
  pages={2748--2759},
  year={2024},
  publisher={Taylor \& Francis}
}

@article{song2024general,
  title={A general m-estimation theory in semi-supervised framework},
  author={Song, Shanshan and Lin, Yuanyuan and Zhou, Yong},
  journal={Journal of the American Statistical Association},
  volume={119},
  number={546},
  pages={1065--1075},
  year={2024},
  publisher={Taylor \& Francis}
}

@article{hoffmann2020consistency,
  title={Consistency of semi-supervised learning algorithms on graphs: Probit and one-hot methods},
  author={Hoffmann, Franca and Hosseini, Bamdad and Ren, Zhi and Stuart, Andrew M},
  journal={Journal of Machine Learning Research},
  volume={21},
  number={186},
  pages={1--55},
  year={2020}
}

@article{wang2007large,
author  = {Junhui Wang and Xiaotong Shen},
  title   = {Large Margin Semi-supervised Learning},
  journal = {Journal of Machine Learning Research},
  year    = {2007},
  volume  = {8},
  number  = {65},
  pages   = {1867--1891}
}

@article{benjamini1995controlling,
  title={Controlling the false discovery rate: a practical and powerful approach to multiple testing},
  author={Benjamini, Yoav and Hochberg, Yosef},
  journal={Journal of the Royal Statistical Society: Series B (Statistical Methodology)},
  volume={57},
  number={1},
  pages={289--300},
  year={1995},
  publisher={Wiley Online Library}
}

@article{wu2021model,
title={Model-assisted uniformly honest inference for optimal treatment regimes in high dimension},
  author={Wu, Yunan and Wang, Lan and Fu, Haoda},
  journal={Journal of the American Statistical Association},
  volume={118},
  number={541},
  pages={305--314},
  year={2023},
  publisher={Taylor \& Francis}
}

@article{bartlett2006convexity,
  title={Convexity, classification, and risk bounds},
  author={Bartlett, Peter L and Jordan, Michael I and McAuliffe, Jon D},
  journal={Journal of the American Statistical Association},
  volume={101},
  number={473},
  pages={138--156},
  year={2006},
  publisher={Taylor \& Francis}
}

@article{ma2020global,
 title={Global and simultaneous hypothesis testing for high-dimensional logistic regression models},
  author={Ma, Rong and Tony Cai, T and Li, Hongzhe},
  journal={Journal of the American Statistical Association},
  volume={116},
  number={534},
  pages={984--998},
  year={2021},
  publisher={Taylor \& Francis}
}

@article{shi2019linear,
  title={Linear hypothesis testing for high dimensional generalized linear models},
  author={Shi, Chengchun and Song, Rui and Chen, Zhao and Li, Runze and others},
  journal={Annals of Statistics},
  volume={47},
  number={5},
  pages={2671--2703},
  year={2019},
  publisher={Institute of Mathematical Statistics}
}

@article{zhang2016variable,
  title={Variable selection for support vector machines in moderately high dimensions},
  author={Zhang, Xiang and Wu, Yichao and Wang, Lan and Li, Runze},
  journal={Journal of the Royal Statistical Society: Series B (Statistical Methodology)},
  volume={78},
  number={1},
  pages={53--76},
  year={2016},
  publisher={Wiley Online Library}
}

@article{zhang2016consistent,
  title={A consistent information criterion for support vector machines in diverging model spaces},
  author={Zhang, Xiang and Wu, Yichao and Wang, Lan and Li, Runze},
  journal={Journal of Machine Learning Research},
  volume={17},
  number={1},
  pages={466--491},
  year={2016},
  publisher={JMLR. org}
}

@article{koo2008bahadur,
author  = {Ja-Yong Koo and Yoonkyung Lee and Yuwon Kim and Changyi Park},
  title   = {A Bahadur Representation of the Linear Support Vector Machine},
  journal = {Journal of Machine Learning Research},
  year    = {2008},
  volume  = {9},
  number  = {44},
  pages   = {1343--1368}
}

@article{zhang2004statistical,
  title={Statistical behavior and consistency of classification methods based on convex risk minimization},
  author={Zhang, Tong},
  journal={Annals of Statistics},
  volume={32},
  number={1},
  pages={56--85},
  year={2004},
  publisher={Institute of Mathematical Statistics}
}

@article{steinwart2005consistency,
  title={Consistency of support vector machines and other regularized kernel classifiers},
  author={Steinwart, Ingo},
  journal={IEEE Transactions on Information Theory},
  volume={51},
  number={1},
  pages={128--142},
  year={2005},
  publisher={IEEE}
}

@article{lin2004note,
  title={A note on margin-based loss functions in classification},
  author={Lin, Yi},
  journal={Statistics \& probability letters},
  volume={68},
  number={1},
  pages={73--82},
  year={2004},
  publisher={Elsevier}
}

@article{blanchard2008statistical,
  title={Statistical performance of support vector machines},
  author={Blanchard, Gilles and Bousquet, Olivier and Massart, Pascal},
  journal={Annals of Statistics},
  volume={36},
  number={2},
  pages={489--531},
  year={2008},
  publisher={Institute of Mathematical Statistics}
}

@article{vert2006consistency,
 author  = {R{{\'e}}gis Vert and Jean-Philippe Vert},
  title   = {Consistency and Convergence Rates of One-Class SVMs and Related Algorithms},
  journal = {Journal of Machine Learning Research},
  year    = {2006},
  volume  = {7},
  number  = {29},
  pages   = {817--854}
}

@article{steinwart2007fast,
  title={Fast rates for support vector machines using Gaussian kernels},
  author={Steinwart, Ingo and Scovel, Clint and others},
  journal={Annals of Statistics},
  volume={35},
  number={2},
  pages={575--607},
  year={2007},
  publisher={Institute of Mathematical Statistics}
}

@article{cortes1995support,
  title={Support-vector networks},
  author={Cortes, Corinna and Vapnik, Vladimir},
  journal={Machine learning},
  volume={20},
  number={3},
  pages={273--297},
  year={1995},
  publisher={Springer}
}

@article{xue2020multicategory,
  title={Multicategory angle-based learning for estimating optimal dynamic treatment regimes with censored data},
  author={Xue, Fei and Zhang, Yanqing and Zhou, Wenzhuo and Fu, Haoda and Qu, Annie},
  journal={Journal of the American Statistical Association},
  volume={117},
  number={539},
  pages={1438--1451},
  year={2022},
  publisher={Taylor \& Francis}
}

@article{pan2020improved,
 title={Improved doubly robust estimation in learning optimal individualized treatment rules},
  author={Pan, Yinghao and Zhao, Ying-Qi},
  journal={Journal of the American Statistical Association},
  volume={116},
  number={533},
  pages={283--294},
  year={2021},
  publisher={Taylor \& Francis}
}

@article{zhao2019efficient,
title={Efficient augmentation and relaxation learning for individualized treatment rules using observational data},
  author={Zhao, Ying-Qi and Laber, Eric B and Ning, Yang and Saha, Sumona and Sands, Bruce E},
  journal={Journal of Machine Learning Research},
  volume={20},
  number={1},
  pages={1821--1843},
  year={2019},
  publisher={JMLR. org}
}

@article{chen2016personalized,
  title={Personalized dose finding using outcome weighted learning},
  author={Chen, Guanhua and Zeng, Donglin and Kosorok, Michael R},
  journal={Journal of the American Statistical Association},
  volume={111},
  number={516},
  pages={1509--1521},
  year={2016},
  publisher={Taylor \& Francis}
}

@article{peng2016error,
  title={An error bound for l1-norm support vector machine coefficients in ultra-high dimension},
  author={Peng, Bo and Wang, Lan and Wu, Yichao},
  journal={Journal of Machine Learning Research},
  volume={17},
  number={1},
  pages={8279--8304},
  year={2016},
  publisher={JMLR. org}
}

@article{bartlett2008classification,
  title={Classification with a reject option using a hinge loss},
  author={Bartlett, Peter L and Wegkamp, Marten H},
  journal={Journal of Machine Learning Research},
  volume={9},
  number={8},
  pages={1823--1840},
  year={2008}
}

@article{wang2019distributed,
  title={Distributed Inference for Linear Support Vector Machine.},
  author={Wang, Xiaozhou and Yang, Zhuoyi and Chen, Xi and Liu, Weidong},
  journal={Journal of Machine Learning Research},
  volume={20},
  number={113},
  pages={1--41},
  year={2019}
}

@article{chernozhukov2017double,
  title={Double/debiased/neyman machine learning of treatment effects},
  author={Chernozhukov, Victor and Chetverikov, Denis and Demirer, Mert and Duflo, Esther and Hansen, Christian and Newey, Whitney},
  journal={American Economic Review},
  volume={107},
  number={5},
  pages={261--65},
  year={2017}
}

@book{imbens_rubin_2015, place={Cambridge}, title={Causal Inference for Statistics, Social, and Biomedical Sciences: An Introduction}, publisher={Cambridge University Press}, author={Imbens, Guido W. and Rubin, Donald B.}, year={2015}}

@article{chunhui2017,
	Author = {Dezeure, Ruben and B{\"u}hlmann, Peter and Zhang, Cun-Hui},
	Da = {2017/12/01},
	Id = {Dezeure2017},
	Isbn = {1863-8260},
	Journal = {TEST},
	Number = {4},
	Pages = {685--719},
	Title = {High-dimensional simultaneous inference with the bootstrap},
	Ty = {JOUR},
	Volume = {26},
	Year = {2017}}

@article{yingqi2014,
	author={Zhao, Ying-Qi and Zeng, Donglin and Laber, Eric B and Song, Rui and Yuan, Ming and Kosorok, Michael Rene},
	title = "{Doubly robust learning for estimating individualized treatment with censored data}",
	journal = {Biometrika},
	volume = {102},
	number = {1},
	pages = {151-168},
	year = {2014},
	eprint = {https://academic.oup.com/biomet/article-pdf/102/1/151/618081/asu050.pdf},
}

@article{jeng2018,
	author = "Jeng, X. Jessie and Lu, Wenbin and Peng, Huimin",
	journal = "Electronic Journal of Statistics",
	number = "1",
	pages = "2074--2089",
	publisher = "The Institute of Mathematical Statistics and the Bernoulli Society",
	title = "High-dimensional inference for personalized treatment decision",
	volume = "12",
	year = "2018"
}

@article{rubin2005,
	author = {Donald B Rubin},
	title = {Causal Inference Using Potential Outcomes},
	journal = {Journal of the American Statistical Association},
	volume = {100},
	number = {469},
	pages = {322-331},
	year  = {2005},
	publisher = {Taylor & Francis}
}

@article{rubin1974,
	title={Estimating causal effects of treatments in randomized and nonrandomized studies.},
	author={Rubin, Donald B},
	journal={Journal of Educational Psychology},
	volume={66},
	number={5},
	pages={688},
	year={1974},
	publisher={American Psychological Association}
}

@article{victor2018,
	author = {Chernozhukov, Victor and Chetverikov, Denis and Demirer, Mert and Duflo, Esther and Hansen, Christian and Newey, Whitney and Robins, James},
	title = {Double/debiased machine learning for treatment and structural parameters},
	journal = {The Econometrics Journal},
	volume = {21},
	number = {1},
	pages = {C1-C68},
	eprint = {https://onlinelibrary.wiley.com/doi/pdf/10.1111/ectj.12097},
	year = {2018}
}

@article{newey1997,
	title = "Convergence rates and asymptotic normality for series estimators",
	journal = "Journal of Econometrics",
	volume = "79",
	number = "1",
	pages = "147-168",
	year = "1997",
	author = "Whitney K. Newey"
}

@article{vandegeer2014,
  title={On asymptotically optimal confidence regions and tests for high-dimensional models},
  author={Van de Geer, Sara and B{\"u}hlmann, Peter and Ritov, Ya’acov and Dezeure, Ruben},
  journal={Annals of Statistics},
  volume={42},
  number={3},
  pages={1166--1202},
  year={2014},
  publisher={Institute of Mathematical Statistics}
}

@article{ning2017,
	author = "Ning, Yang and Liu, Han",
	journal = "Annals of Statistics",
	number = "1",
	pages = "158--195",
	publisher = "The Institute of Mathematical Statistics",
	title = "A general theory of hypothesis tests and confidence regions for sparse high dimensional models",
	volume = "45",
	year = "2017"
}

@article{Zhao2012,
	author = {Zhao, Yingqi and Zeng, Donglin and Rush, A John and Kosorok, Michael R},
	journal = {Journal of the American Statistical Association},
	number = {499},
	pages = {1106--1118},
	title = {{Estimating individualized treatment rules using outcome weighted learning}},
	volume = {107},
	year = {2012}
}

@article{Zhou2017,
author = {Xin Zhou and Nicole Mayer-Hamblett and Umer Khan and Michael R. Kosorok},
title = {Residual Weighted Learning for Estimating Individualized Treatment Rules},
journal = {Journal of the American Statistical Association},
volume = {112},
number = {517},
pages = {169-187},
year  = {2017},
publisher = {Taylor & Francis}
}
\end{document}